\documentclass[twocolumn]{aastex631}
\usepackage{color}
\usepackage{amsmath}

\newcommand{\halpha}{H{$\alpha$}}

   \font\sevenrm=cmr7 scaled 1000

\newcommand{\SIII}{[S{\sevenrm III}]}

\usepackage{soul}
\usepackage{changepage}

\begin{document}

\title{NEXUS: A Search for Nuclear Variability with the First Two JWST NIRCam Epochs}

\author[0000-0002-8501-3518]{Zachary Stone}
\affiliation{Department of Astronomy, University of Illinois at Urbana-Champaign, Urbana, IL 61801, USA}

\author[0000-0003-1659-7035]{Yue Shen}
%\email{shenyue@illinois.edu}
\affiliation{Department of Astronomy, University of Illinois at Urbana-Champaign, Urbana, IL 61801, USA}
\affiliation{National Center for Supercomputing Applications, University of Illinois at Urbana-Champaign, Urbana, IL 61801, USA}

\author[0000-0001-5105-2837]{Ming-Yang Zhuang}
%\email{mingyang@illinois.edu}
\affiliation{Department of Astronomy, University of Illinois at Urbana-Champaign, Urbana, IL 61801, USA}

\author[0000-0001-7201-1938]{Lei Hu}
\affiliation{McWilliams Center for Cosmology, Department of Physics, 
Carnegie Mellon University, 5000 Forbes Ave, Pittsburgh, 15213, PA, USA}

\author[0000-0002-2361-7201]{Justin Pierel}
%\email{mingyang@illinois.edu}
\affiliation{Space Telescope Science Institute, Baltimore, MD 21218, USA}
\altaffiliation{NASA Einstein Fellow}

\author[0000-0002-1605-915X]{Junyao Li}
\affiliation{Department of Astronomy, University of Illinois at Urbana-Champaign, Urbana, IL 61801, USA}

%%%%%%%%%%%%%%%%%%

\author[0000-0002-6523-9536]{Adam J.\ Burgasser}
\affiliation{Department of Astronomy \& Astrophysics, UC San Diego, La Jolla, CA 92093, USA}

\author[0000-0002-5612-3427]{Jenny E. Greene}
\affil{Department of Astrophysical Sciences, 4 Ivy Lane, Princeton University, Princeton, NJ 08540}

\author[0000-0003-0230-6436]{Zhiwei Pan}
\affiliation{Department of Astronomy, University of Illinois at Urbana-Champaign, Urbana, IL 61801, USA}

\author[0000-0003-3509-4855]{Alice E. Shapley}
\affiliation{Department of Physics \& Astronomy, University of California, Los Angeles, 430 Portola Plaza, Los Angeles, CA 90095, USA}

\author[0000-0002-4622-6617]{Fengwu Sun}
\affiliation{Center for Astrophysics $|$ Harvard \& Smithsonian, 60 Garden St., Cambridge, MA 02138, USA}

\author[0000-0001-8638-2780]{Padmavathi Venkatraman}
\affiliation{Department of Astronomy, University of Illinois at Urbana-Champaign, Urbana, IL 61801, USA}

\author[0000-0002-7633-431X]{Feige Wang}
\affiliation{Department of Astronomy, University of Michigan, 1085 S. University Ave., Ann Arbor, MI 48109, USA}

%\author[0000-0001-6052-4234]{Xiaojing Lin}
%\affiliation{Department of Astronomy, Tsinghua University, Beijing 100084, China}
%\affiliation{Steward Observatory, University of Arizona, 933 N Cherry Ave, Tucson, AZ 85721, USA}

%\author[0000-0002-5612-3427]{Jenny E.~Greene}
%\affiliation{Department of Astrophysical Sciences, Princeton University, Princeton, NJ 08544, USA}

%\author[0000-0003-3509-4855]{Alice E.~Shapley}
%\affiliation{Department of Physics \& Astronomy, University of California, Los Angeles, 430 Portola Plaza, Los Angeles, CA 90095, USA}

%\author[0000-0002-7633-431X]{Feige Wang}
%\affiliation{Department of Astronomy, University of Michigan, 1085 S. University Ave., Ann Arbor, MI 48109, USA}

%\author[0000-0003-4202-1232]{Qiaoya Wu}
%\affiliation{Department of Astronomy, University of Illinois Urbana-Champaign, Urbana, IL 61801, USA}

%\author[0000-0002-6893-3742]{Qian Yang}
%\affiliation{Center for Astrophysics $\vert$ Harvard \& Smithsonian, 60 Garden Street, Cambridge, MA 02138, USA}

%\author{et al.}

\begin{abstract}
The multicycle JWST Treasury program NEXUS will obtain cadenced imaging and spectroscopic observations around the North Ecliptic Pole during 2024--2028. Here we report a systematic search for nuclear variability among $\sim 25,000$ sources covered by NIRCam (F200W+F444W) imaging using the first two NEXUS epochs separated by 9 months in the observed frame. Difference imaging techniques reach $1\sigma$ variability sensitivity of 0.18~mag (F200W) and 0.15~mag (F444W) at 28th magnitude (within 0\farcs2 diameter aperture), improved to $0.01$~mag and $0.02$~mag at $<25$th magnitude, demonstrating the superb performance of NIRCam photometry. The difference imaging results represent significant improvement over aperture photometry on individual epochs (by $>30\%$). We identify 465 high-confidence variable sources among the parent sample, with a two-epoch flux difference at $>3\sigma$ from the fiducial variability sensitivity. Essentially, all these variable sources are of extragalactic origin based on preliminary photometric classifications, and follow a similar photometric redshift distribution as the parent sample up to $z_{\rm phot}>10$. While the majority of these variability candidates are likely normal unobscured active galactic nuclei, some of them may be rare nuclear stellar transients and tidal disruption events that await confirmation with spectroscopy and continued photometric monitoring. We also constrain the photometric variability of 10 spectroscopically confirmed broad-line Little Red Dots (LRDs) at $3\lesssim z \lesssim 7$, and find none of them show detectable variability in either band. We derive stringent $3\sigma$ upper limits on the F444W variability of $\sim 3-10\%$ for these LRDs, with a median value of $\sim 5\%$. These constraints imply weak variability in the rest-frame optical continuum of LRDs.  
\end{abstract}

%We find substantial contamination of photometric LRD selection based on six-band NIRCam imaging from NEXUS, highlighting the necessity of spectral confirmation for sample purity and completeness.

%\blue{(6/13 broad line LRDs)}

%% Keywords should appear after the \end{abstract} command. 
%% The AAS Journals now uses Unified Astronomy Thesaurus concepts:
%% https://astrothesaurus.org
%% You will be asked to selected these concepts during the submission process
%% but this old "keyword" functionality is maintained in case authors want
%% to include these concepts in their preprints.

%Brown dwarfs (185), Supernovae (1668), 

\section{Introduction} \label{sec:intro}

Recent deep multi-epoch JWST imaging observations have enabled a new avenue for the identification and characterization of faint transients and variables in the distant Universe \citep{Yan_2023, DeCourseyEtAl2025}. With the superb sensitivity at infrared wavelengths, these JWST observations can not only reveal stellar explosions far beyond the reach of ground telescopes and the Hubble Space Telescope (HST), but they also provide an efficient means for detecting tenuous nuclear variability that may signal the presence of an accreting supermassive black hole (SMBH). Dedicated JWST survey programs in several well-studied fields further provide high-quality source catalogs with well-characterized photometric properties, as well as spectroscopy, to facilitate the study of transients and variables discovered in these legacy fields \citep[e.g.,][]{JADES,PEARLS,nexus}. 

At the JWST wavelengths and depths, transient detection appears abundant, e.g., $\sim$ a few per ${\rm arcmin^2}$ per year at a $\sim 30$ magnitude limit \citep[e.g.,][]{DeCourseyEtAl2025}, and reaches redshifts of $z\gtrsim 3$ for the first time \citep[e.g.,][]{Yan_2023,DeCourseyEtAl2025,DeCourseyEtAl2025a}. On the other hand, the sensitivity of the JWST imaging is also crucial for detecting active galactic nucleus (AGN) variability in low-mass, high-redshift systems \citep[e.g.,][]{ZhangZJ2024}, where their faint magnitudes have eluded detection by non-JWST facilities. In particular, an elusive population of faint (e.g., $m_{\rm F444W}\sim 23-27$) broad-line emitters with red rest-optical color and compact morphology in the rest-frame optical, dubbed ``Little Red Dots'' (LRDs), have been identified from early JWST observations \citep[e.g.,][]{Matthee+24,Greene+2024, Kocevski_LRD_selection, Labbe+2025ApJ, Hainline+2025_LRD_selection}. The nature of these LRDs is still being debated, and nuclear variability detection can corroborate the scenario of accreting SMBHs that produce the broad emission lines, and provide useful constraints on accretion models \citep[e.g.,][]{Kokubo2024,SecundaEtAl2025,Zhou_etal_2025}.    

There are several ongoing programs with repeated JWST imaging and spectroscopic observations in deep extragalactic fields \citep[e.g.,][]{JADES,PEARLS,COSMOS-Web,nexus, SAPPHIRES}, with a science goal of detecting transients and variables that can only be reached with JWST. Among these programs, the North ecliptic pole EXtragalactic Unified Survey (NEXUS) multicycle JWST treasury program \citep{nexus} is specifically designed to enable systematic investigations of the variable sky in the JWST era. NEXUS targets a field around the North Ecliptic Pole, which is within the Continuous Viewing Zone of L2 space facilities. This field selection provides unconstrained visibility, and the greatest scheduling flexibility for cadenced monitoring and follow-up observations. In contrast, most of the well-studied extragalactic fields can only be observed by JWST in limited annual visibility windows, preventing regular monitoring programs and flexible follow-up of high-value transient and variable targets. The NEXUS program has two overlapping tiers, with the Wide area covering $\sim 0.1\,{\rm deg^2}$ with three annual epochs, and the Deep area covering the central $\sim 50\,{\rm arcmin^2}$ with 18 epochs over 2025-2028 with a fixed 2-month cadence. Each epoch will obtain JWST imaging (NIRCam primary) and spectroscopy (either with NIRCam/WFSS for Wide or with NIRSpec/MSA/PRISM for Deep). The NEXUS survey obtained its first (partial) Wide epoch in September 2024 \citep{nexus-edr}, covering the Deep area with reference NIRCam images. The first NEXUS-Deep epoch was obtained in June 2025. An approximate timeline of scheduled future epochs through 2028 is presented in the NEXUS overview paper \citep{nexus}. 

In this work we perform an initial study of the nuclear variable population in NEXUS using the first Wide epoch and the first Deep epoch with an overlapping area of $\sim 40\,{\rm arcmin^2}$, separated by 9 months in the observed frame. The main purposes of this study are to quantify the nuclear (i.e., $<0\farcs1$ from the source centroid) variability properties of NEXUS sources detected by NIRCam at $m_{\rm F444W}\lesssim 28$ and identify those with nuclear variability, and to investigate the variability properties of spectroscopically-confirmed LRDs therein \citep{ZhuangEtAl2025a}. Off-nucleus transients detected with these NEXUS epochs are reported elsewhere. In Section~\ref{sec:data} we describe the data used for variability detection. We describe in detail the methodology for detecting variable sources using multi-epoch JWST imaging in Section~\ref{sec:methods} and present the nuclear variability source catalog in Section~\ref{sec:results}. We discuss our results in Section~\ref{sec:disc} and conclude in Section~\ref{sec:con}.

%\red{Tasks for Zach: 1. perform difference imaging with Wide-1 and Deep-1 and obtain magnitude changes for all sources -- test SFFT and PyZogy. 2. identify candidate nuclear variability sources, and discuss their nature using the compiled NEXUS source catalog. 3. discuss the variability (or non-variability) of covered spectroscopic and photometric LRDs. }\red{Make sure you obtain consistent variability or difference fluxes regardless which difference imaging method you use. }

\section{Data}\label{sec:data}

%\red{Describe the Wide-1 and Deep-1 data, and the photometric source catalog. Describe astrometry alignment [Mingyang]. }

The first NIRCam imaging epoch was carried out on {September 12-13}, 2024 as part of the NEXUS-Wide Epoch 1. It covers the central $\sim 100\,{\rm arcmin^2}$ area with six-band NIRCam imaging (F090W, F115W, F150W, F200W, F356W, F444W). The typical $5\sigma$ depth (AUTO mag) is $m_{\rm F444W}\sim 28$. More details about these observations can be found in \citet{nexus-edr}. We refer to this epoch as Wide-1.1. The second NIRCam imaging epoch was carried out during June 1-2, 2025 as part of the NEXUS-Deep Epoch 1 that performs NIRSpec MSA spectroscopy and NIRCam/MIRI imaging over the central NEXUS area. Here, we only use the primary NIRCam imaging associated with Deep Epoch 1, which only covers F200W and F444W with a total area of $\sim70$ arcmin$^2$ and a $5\sigma$ depth $m_{\rm F444W}\sim 27.5$ (AUTO mag). We refer to this epoch as Deep-1. Although the final coadded data from all NEXUS-Deep epoch imaging will be deeper than the coadded WIDE data, the depth of an individual Deep epoch is shallower. The Wide-1.1 epoch had a total exposure time of 1234 s for the majority of the F200W coverage and 934.1 s for F444W, while the Deep-1 epoch had 311.4 s for approximately half of the coverage with one exposure and 622.7 s for the remaining part.

\subsection{Processing the Two-Epoch Images}

\begin{figure}[!t]
\centering
  \includegraphics[width=0.48\textwidth]{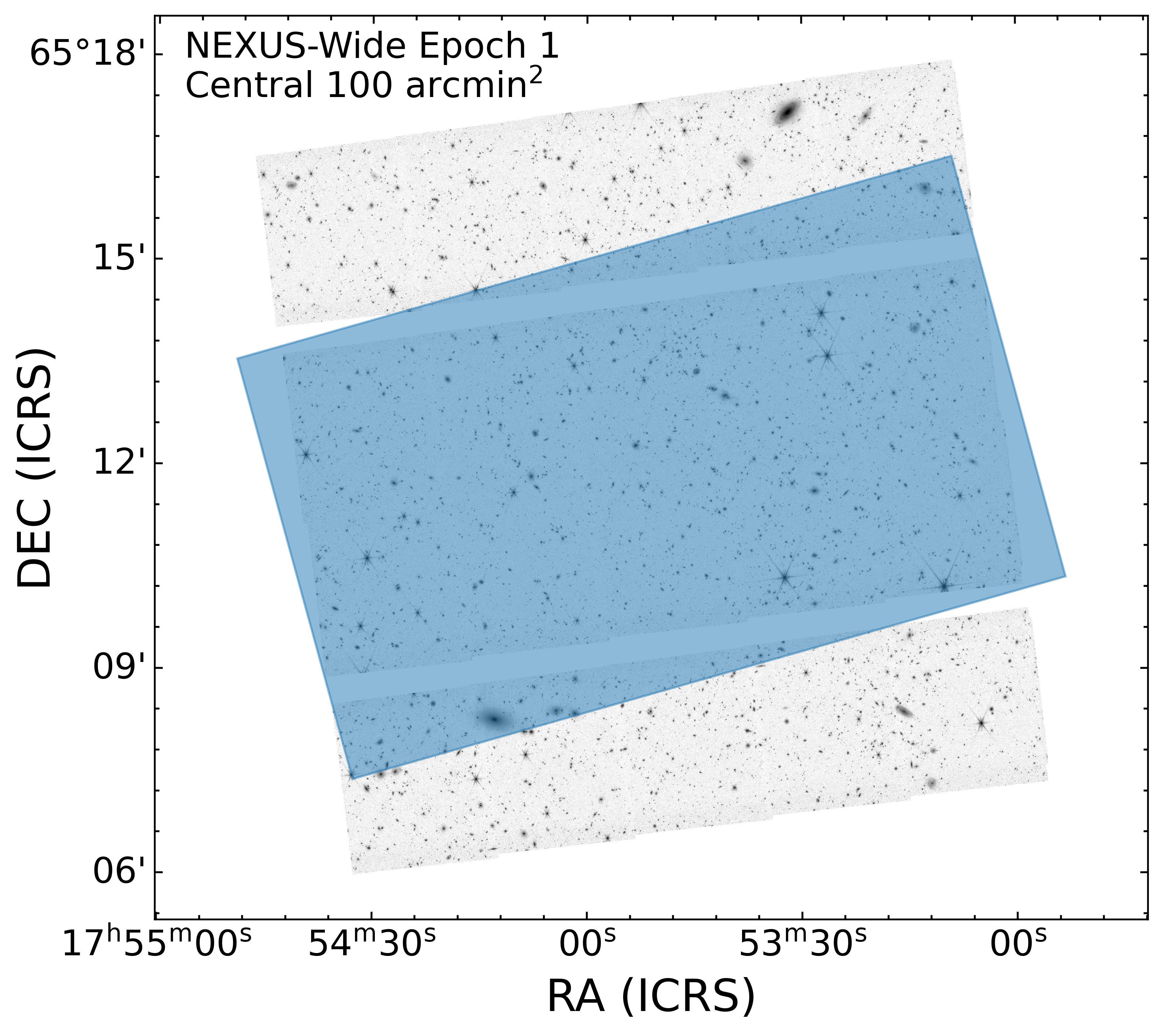}
  \caption{NIRCam imaging overlap between the first (partial) Wide epoch (Wide-1.1; background image) and the first Deep epoch (Deep-1; shaded blue). 
  \label{fig:layout}}
\end{figure}

We perform NIRCam imaging reduction for the new data from the Deep-1 epoch following the procedures for the first NIRCam epoch (Wide-1.1) data presented in the NEXUS Early Data Release \citep[EDR;][]{nexus-edr}. The standard reduction steps include stage 1 detector-level corrections, stage 2 image calibration, and stage 3 mosaic construction. Prior to stage 3, we remove a pedestal background, ``1/f'' noise, and large-scale ``wisp'' emission, and mask other artifacts caused by scattered light\footnote{\url{https://jwst-docs.stsci.edu/known-issues-with-jwst-data/nircam-known-issues/nircam-scattered-light-artifacts}, including dragon's breath, large-scale parallel striping, and ``claws''}. We aggressively mask the source emission during ``1/f'' noise removal step and use a high threshold for per-channel noise removal to achieve better performance. In the meantime, we adopt the input pixel “shrunk” fraction \texttt{pixfrac} of 1 instead of 0.8 for the NEXUS EDR in the poorly sampled images (i.e., the Deep epochs) with only 1 or 2 dithered exposures. We further refine the astrometric alignment by: (1) grouping long-wavelength (LW) detector images of the same pointings together as they share the same field of view; and (2) iteratively grouping different pointings together if their overlapping area $>1$ arcmin$^2$ to secure robust relative astrometry among each detector image. We finally tie the internally aligned detector image groups to GAIA DR3 if at least 15 objects are matched, otherwise to the Subaru Hyper Suprime-Cam (HSC) catalog of the Hawaii eROSITA Ecliptic Pole Survey \citep[HEROES;][]{HEROES}. Therefore, we re-reduce the first epoch data to ensure consistency among different epochs. 
%The reduced images from the prior (central) Wide Epoch 1 are taken from NEXUS EDR \citep{nexus-edr}. 
For each epoch, we merge the dithered images to produce a single-epoch mosaic. Our final mosaics have a pixel scale of 0\farcs03, with an additional 0\farcs06 version for the F444W filter. The observing time differences during each epoch ($\sim$5 min.) across the dithered images are negligible for the interested variability timescales (months) in this work. The overlays of NIRCam images from these two epochs are shown in Fig.~\ref{fig:layout}. There is approximately a $\sim 60\,{\rm arcmin^2}$ area covered by both epochs, which is the primary area for our variability study.

\subsection{The Parent Source Catalog}\label{subsec:catalog}

In order to examine source variability, we first build a parent source catalog from the stacked two-epoch images and compile source properties in Table~\ref{tab:catalog}. The locations of the parent sources are later used to extract fluxes in individual epochs and in the difference images for our variability analyses. First, we smooth the inverse-variance weight maps to remove remaining spurious sources and bad pixels. Bright stars, including their diffraction spikes, are then masked using a magnitude-dependent threshold. The detection image is built by coadding inverse-variance-weighted mosaics in the F200W, F210M, F356W, F360M, and F444W filters in both epochs, with F210M and F360M from parallel exposures of primary NIRSpec MOS observations. Following the NEXUS EDR, we run \texttt{Source Extractor} \citep[\texttt{SExtractor;}][]{sextractor} in dual-mode on the overlapping area between the two epochs. We choose a relative detection and analysis threshold of 1.5$\sigma$, a deblending threshold of 32, and a minimum contrast parameter of 0.001. It should be noted, given the different detector coverages in the short and long channels of NIRCam, the overlap between the two epochs is not completely covered by all filters. 

%\red{Zach, please check sources near the image edges -- these are potentially sources that would have large systematic uncertainties in the differential fluxes. }

There are in total 24,875 sources identified from the stacked NIRCam image, although not all sources are detected in individual epochs. In addition, not all sources in the parent catalog have F200W+F444W coverage. We utilize the full source catalog, as all sources therein have at least one band (F200W or F444W) covered in both epochs such that variability measurements in that band are possible. A subset of 14,583 sources is further covered by both epochs and both bands. 

%\red{[Mingyang] Were all sources reported in the EDR source catalog? If not, why? } \blue{Should be some not reported in the EDR catalog, including fainter objects and deblended objects. This is because (1) my new detection image includes more filters (deeper); (2) the inclusion of short wavelength filters would lead to better resolution (better deblending); (3) the use of a smaller deblending contrast (0.001 versus 0.005). }

%\red{Mingyang and Zach, please make sure that the parent catalog source numbers are consistent across this paper. My understanding is that for each parent catalog source, there should be at least one measured epoch flux per band per epoch. However, some sources may have near-zero epoch flux in a specific band and/or epoch, i.e., consistent with background flux. You need to explain how you treat such sources when you compute $\Delta m$ below. }

Given the positions of these sources, we extract aperture fluxes in individual epochs, regardless of the detection significance. We adopt different aperture sizes (e.g., AUTO, 0\farcs2, 0\farcs3, 0\farcs5 diameter), with the AUTO aperture capturing most of the flux from the source at the cost of enclosing more background pixels than smaller apertures. We will compare the differences between the epoch photometry and the variable fluxes measured from difference images in Section~\ref{sec:methods}. 

The NIRCam images are astrometrically aligned such that the typical rms in relative astrometry is $8$~mas and $9$~mas in RA and DEC respectively across different filters and epochs. This means for sources where the centroid is physically well defined, even our smallest $0\farcs2$-diameter aperture will capture variable fluxes within 0\farcs1-radius from the source centroid.  

%\red{[Mingyang] Here we should comment on the accuracy of the source centroid in the parent catalog. Uncertainties come from astrometric errors, asymmetric extended sources, close companions, etc. It would be good to quote a typical absolute uncertainty in the source centroid for individual sources, e.g., 0\farcs02? Note this is NOT the same as the mean astrometric offset derived from the reference Gaia catalog. } \blue{I think we can only get an estimate of relative astrometry errors, which is 8 mas in RA and 9 mas in DEC among different filters and epochs. It is very difficult to quantify the error introduced by asymmetric emission and close companions as they would depend on specific cases. May combine the half-light radius with the relative accuracy to propagate the uncertainty in the flux: consider the median/average size of the sources, obtain how much difference in flux would be caused due to a random shift in position (2D elliptical gaussian). For source profile, use PSF for point-like source and use sersic index=1, 2, 4 for typical galaxies. Do it for three apertures: r=0.2, 0.3 and 0.5.}

\section{Variability Measurements}\label{sec:methods}

%\red{Define nuclear variability: separation from centroid $<0\farcs5$? Point source detected with DAOStarFinder in the difference image. Only one-band detection would suffice (mention chance artifacts -- must be extremely unlucky to land on the source centroid). }

Difference imaging provides a more robust way to identify variable sources than using individual-epoch photometry, especially for extended sources and the search for tenuous nuclear variability. In this work we test two difference imaging methods, in comparison to individual-epoch photometry. The first one is based on the direct subtraction of frame-matched two-epoch images. Given the superb stability and astrometric alignment of NIRCam imaging, direct subtraction often provides sufficient quality to identify variable fluxes, and is commonly adopted in transient searches with JWST data \citep[e.g.,][]{Yan_2023, DeCourseyEtAl2025}. The second is based on the Saccadic Fast Fourier Transform (SFFT) difference imaging algorithm \citep{HuEtAl2022a, HuWang2024a} specifically designed for JWST images. SFFT has the ability to match the point-spread function (PSF) between two images and model spatial variations in the PSF over each image, as well as the differential background in the difference image and spatial variations in photometric scaling. Furthermore, SFFT performs image subtraction in the Fourier domain, allowing for fast, highly parallelized computation.

\begin{figure*}
    \centering
    \includegraphics[width=\linewidth]{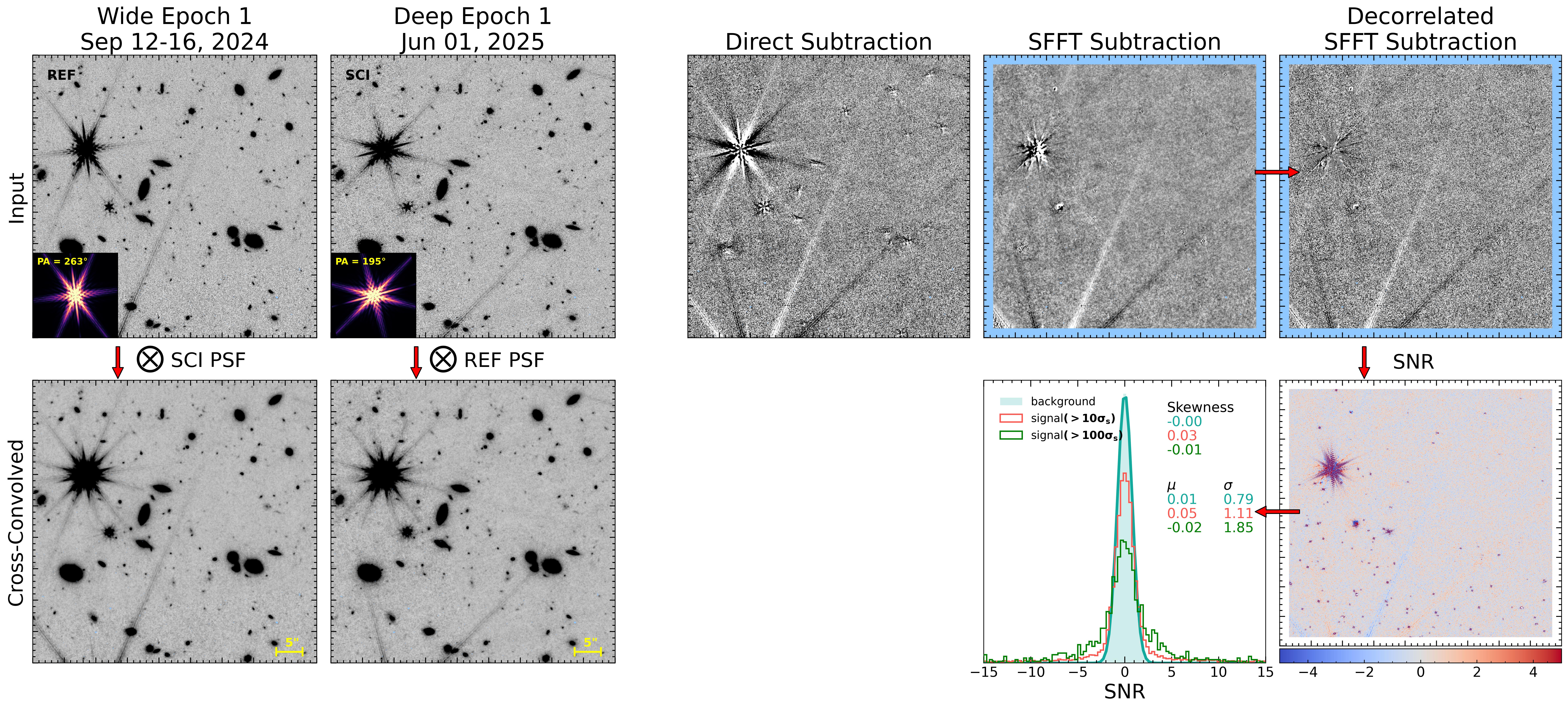}
    \caption{The procedure used to generate our difference images, displaying an example 900$\times$900 pixel cutout in the F444W band. The sky-subtracted cutouts are shown in the two upper left panels, and the cross-convolved cutouts are shown directly beneath them. The PSFs used for each of the input images are shown in insets in the two upper left panels, along with their position angles. In the top right, three panels show the difference images resulting from direct subtraction, SFFT, and SFFT after decorrelation (the latter being the final SFFT results). The lower right panel contains the SNR map resulting from the decorrelated SFFT image, and the distribution of SNR across the map is shown in the panel directly to its left. The SNR of each pixel is mapped to a color defined by the color bar beneath the SNR map. The blue histogram represents background pixels, while the (red, green) histograms represent pixels $>$(10,100)$\sigma$ above the background. The solid blue line is a Gaussian fit to the background pixel distribution. The mean $\mu$, standard deviation $\sigma$, and skewness of these distributions are displayed on the right half of this panel. Masked pixels are shown in light blue in each of the images, except for the SNR map, where they are shown in white.}
    \label{fig:subtraction_quality}
\end{figure*}

Once the difference images are produced, we measure aperture fluxes centered at the source centroid locations in the parent catalog. As for the individual epoch-photometry case, we adopt different aperture sizes. For the purpose of identifying nuclear variability, we use the 0\farcs2 aperture as the default, which provides the best sensitivity, given the smaller background area included compared with larger apertures. Since the variable flux should be point-like, we can scale the aperture flux to the PSF flux with a constant scaling factor. Variable sources that are blended with the parent source in the stacked image and lie much farther away from the nucleus than $0\farcs1$ will be excluded from our variability search with the fiducial 0\farcs2-diameter aperture. On the other hand, isolated transient sources will be detected in the stacked image and hence will have been included in our parent source catalog as separate sources. 

%\blue{We measure the position of the potential variable flux using the brightest pixel within 0\farcs5 of the input centroid position (background subtracted) to quantify the nuclear separation. }

%\red{Some plots needed: 1. examples of detection image of nuclear variability in 2 bands or only 1 band, along with their variability SED. 2. a variability (in mag) versus first-epoch mag plot for all sources, with LRDs and BLAGNs highlighted -- also show histogram plots. }

We convert the linear fluxes measured from the difference image (i.e., variable fluxes) to magnitude units following the equation below:
\begin{equation}
    \Delta m\equiv -2.5\log\left(\frac{f_2}{f_1}\right)=-2.5\log\left(1+\frac{\Delta f}{f_1}\right)\ ,
\end{equation}
where $\Delta f\equiv f_2-f_1$ is the linear differential flux between the two epochs. In this convention, a negative $\Delta m$ or positive $\Delta f$ corresponds to source brightening. To differentiate the magnitude changes under different methodologies (i.e., epoch photometry, direct subtraction, and SFFT), we add the subscripts ${\rm _{epoch}}$, ${\rm _{dir-sub}}$, and ${\rm _{SFFT}}$ to $\Delta m$. 

%\red{Zach should finish the rest of Sec 3.}

\subsection{Direct Frame Subtraction}

%\red{Describe basic procedure and point to example figures. }

We first produce difference images between the two NEXUS epochs via direct subtraction. We prepare the input mosaics in the same manner for both direct subtraction and SFFT for consistency. First, we create a mask, where pixels masked in either of the mosaics are masked in both mosaics. This helps in the SFFT subtraction, where the differential background is modeled, and could create artifacts where one mosaic is masked but the other is not. We then crop the edges of the mosaics to remove as many empty pixels as possible, to reduce the computational resources required by SFFT. Finally, we use the zero-points generated by the JWST pipeline to convert the fluxes in all mosaics to $\mu$Jy. This aids in both the detection of sources in the difference images and the modeling performed in SFFT, as if the pixel values are too low, complications could arise from truncation error. The difference image is then obtained by using the Wide Epoch as the reference image, and the Deep Epoch as the science image.

\subsection{Difference Imaging with SFFT}

%\red{Describe basic procedure and point to example figures. Make sure to mention details, parameters used in SFFT, and flux scaling factors, etc. }

Next, we perform image subtraction between the two NEXUS epochs with SFFT. We generally follow the same procedure described in \citet{HuWang2024a}, and include a few necessary updates. For more information on the implementation and testing of SFFT, see \citet{HuEtAl2022a} and \citet{HuWang2024a}. Firstly, we subtract a constant sky value for each epoch. We use \texttt{NoiseChisel} \citep{noisechisel} to obtain a mask of the background and the SFFT code to obtain the sky value. We then perform cross-convolution between the two sky-subtracted epochs, convolving the Wide Epoch mosaic with the Deep Epoch's PSF and vice versa. Following \citet{HuWang2024a}, we use the \texttt{STPSF} \citep{webbpsf} tool to obtain an approximate PSF for each band, then use \texttt{SWarp} \citep{SWarp} to rotate each PSF to the same orientation as the mosaic in each epoch. While using an empirical PSF would be more accurate for matching the two images, SFFT only requires a rough alignment to guide it toward a well-behaved solution. In many cases, using an empirical PSF will cause overfitting or artifacts. SFFT is fairly computationally expensive, especially with the limited amount of memory native to state-of-the-art GPUs. For this reason, we split the sky-subtracted mosaics into 900$\times$900 pixel cutouts across the whole image -- the same image size as used in \citet{HuWang2024a}. Similarly, we split up the cross-convolved images and error maps.

Next, we construct a mask used by SFFT to perform PSF matching, which masks out ``bad" pixels  (e.g., saturated pixels, variable sources, etc.) and background pixels. The initial parameter selections for mask creation were based on configurations from \citet{nexus-edr}, but were adjusted in order to improve the subtraction for bright and extended sources. The quality of the resulting difference image is sensitive to bright and extended sources, so we construct our maps accordingly. We use \texttt{SExtractor} \citep{sextractor} to identify and extract source information on each cross-convolved cutout separately, using a minimum area of 5 pixels, a detection and analysis threshold of $1\sigma$ above the background, 64 deblending thresholds, and a minimum deblending contrast parameter of $10^{-3}$. A source's pixels are removed from the mask if the source has a \texttt{SExtractor} catalog value FLAG=0 and is either only detectable on one of the mosaics, or has a difference in AUTO magnitude greater than one. Pixels identified by \texttt{SExtractor} as background are also removed from the mask. We then remove all pixels below $3\sigma$, where $\sigma$ is the standard deviation of the background pixels in either of the cross-convolved mosaics. For this standard deviation, we use the minimum between the standard deviation calculated from the whole mosaic and the standard deviation calculated using only the cutout. This helps in cutouts with large background fluxes, say due to a bright star.

\begin{figure*}
    \centering
    \includegraphics[width=\linewidth]{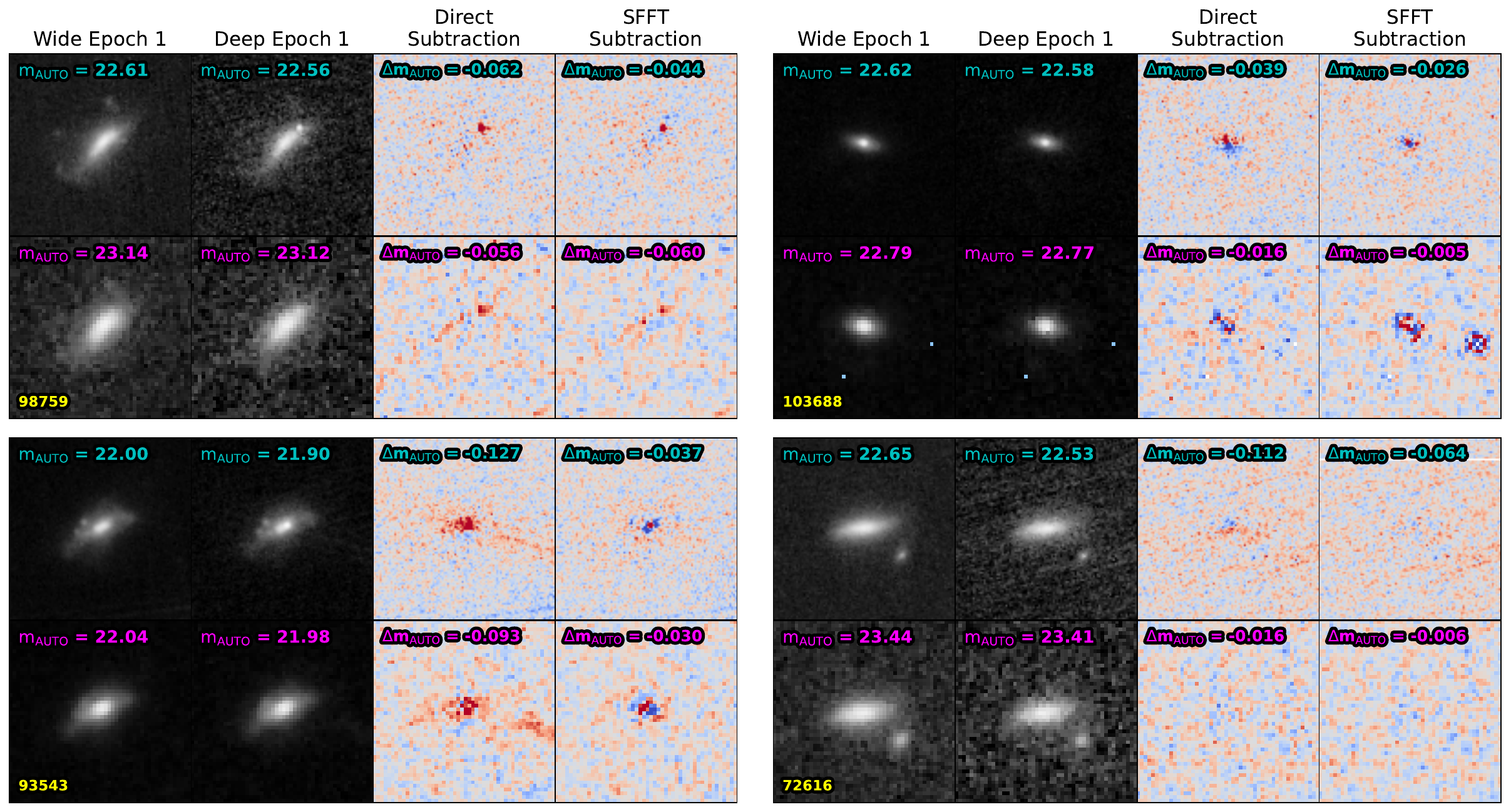}
    \caption{A few example source cutouts from both NEXUS epochs, as well as the difference images from both subtraction methods to compare their performance. For each source, the top (bottom) row represents images in the F200W (F444W) band. The image shown for each column is displayed at the top of each column. The color scale used for the difference images ranges from $\pm 5\sigma$ of the background pixels, with blue representing negative and red positive. For the two NEXUS epochs, the magnitudes of the sources using the AUTO flux are shown in the top left of the panels. For the difference images, the $\Delta m$ of the sources using the AUTO flux are shown in the top left of the panels. Each source's ID from the parent catalog is shown in the bottom left of the lower left panel. Masked pixels are shown in light blue in each of the epoch images and in white in each of the difference images. In general, both subtraction methods perform well, with SFFT performing slightly better in bright cases (further discussed in Section~\ref{sec:comp_diffimg}).}
    \label{fig:example_sources}
\end{figure*}

We then remove all saturated pixels, mainly those associated with bright, saturated stars. We use the global {\tt SExtractor} catalogs to identify bright stars, and the individual cutout {\tt SExtractor} catalogs for dimmer stars. The threshold between bright and dim stars is a fiducial FLUX\_APER = 30 $\mu$Jy using an aperture 5 times the FWHM in a given band. Using the cutout catalogs, we declare sources as stars if they have a {\tt SExtractor} catalog value CLASS\_STAR $> 0.98$, ELONGATION $< 1.3$, and FLUX\_APER $> 1$ $\mu$Jy, using the same aperture of 5$\times$FWHM. Using the global catalogs, we use two sets of conditions, so as to not mask out bright galaxies that cover substantial fractions of the cutouts. Either the source has: 1) 45 $\mu$Jy $<$ FLUX\_APER $< 150$ $\mu$Jy and CLASS\_STAR $>0.9$, or 2) FLUX\_APER $> 150$ $\mu$Jy and CLASS\_STAR $> 0.7$. We have finely tuned this \texttt{SExtractor} configuration with visual inspection of the SFFT masks. Most bright stars and galaxies are masked, with only a few cutouts having poor subtractions because of improper masks. %Generally, this only masks bright stars, but sometimes masks bright, circular galaxies. In these cases, the subtraction will perform worse than if it were not masked, although on average for most sources this suffices.

SFFT is run on each cutout individually, with the same parameters for a given band. We use the Wide Epoch as the reference image, and the Deep Epoch as the science image, choosing to match the PSF of the reference image to that of the science image. For both bands, we use a second-order B-Spline to model spatial kernel variations, a first-order polynomial to model photometric scaling variations, and a constant to model the differential background. We set the kernel half-width to 11 pixels for F200W and 5 pixels for F444W. We set the internal knots for the B-Spline to be located in a 3$\times$3 grid evenly across the whole cutout for F200W, and a 6$\times$6 grid for F444W. We use a Tikhonov regularization parameter $\lambda = 3 \times 10^{-5}$.

Because the cutouts are all of equal size, and the mosaics are rotated with respect to the image plane, some cutouts will have only a small number of non-zero pixels. To prevent overfitting, we use a simpler form to model spatial kernel variations when the number of non-zero pixels is less than 75\% of the total pixels in the cutout. In this case, we use a second-order polynomial to model PSF variations, and keep all other parameters the same.

\begin{figure*}[!t]
    \centering
    \includegraphics[width=\linewidth]{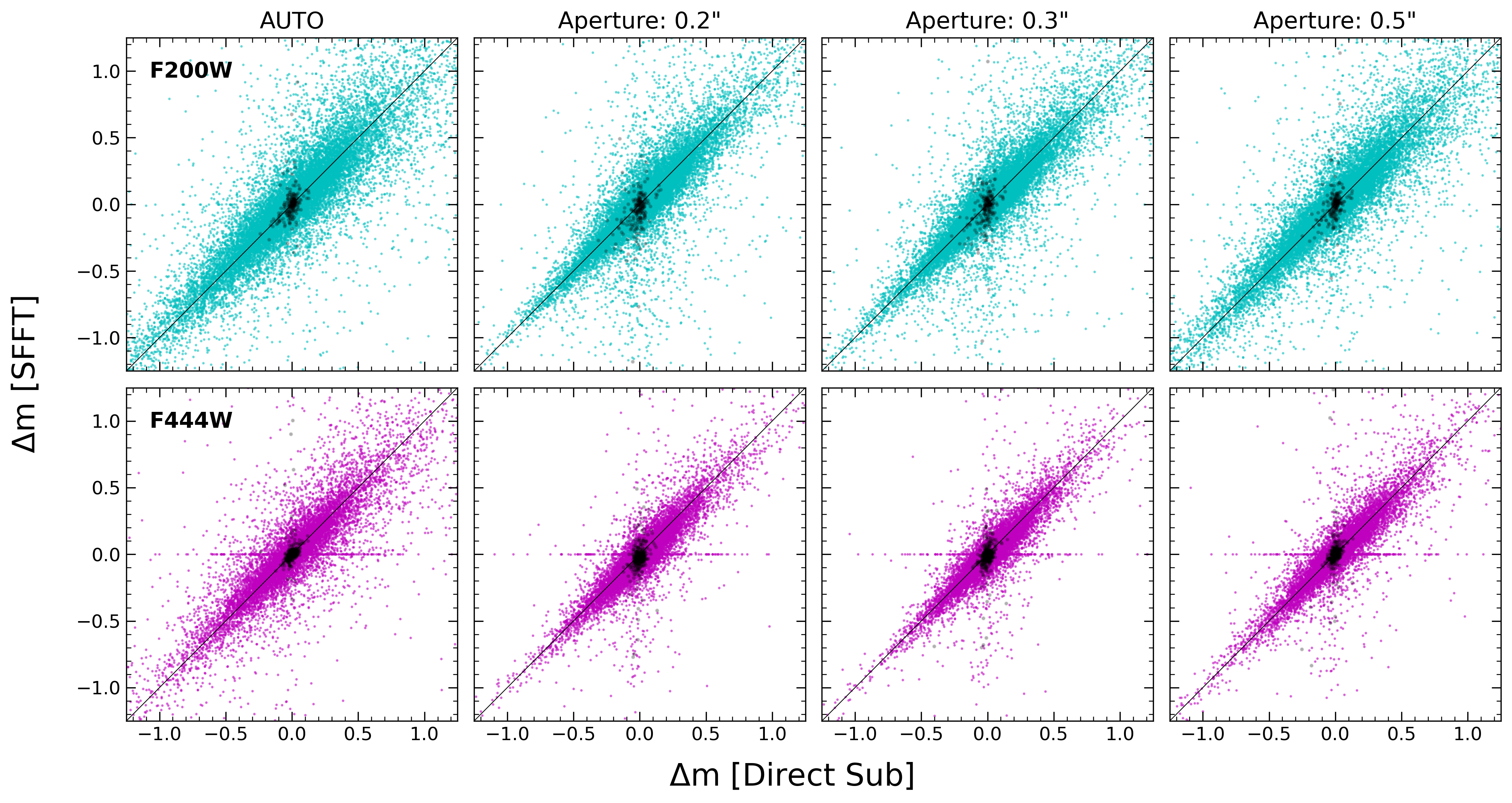}
    \caption{A comparison of the magnitude changes $\Delta m$ of sources identified within the NEXUS field between both subtraction methods, with $m_{\rm DIFF, BEST} < 35$ using a 0\farcs2 aperture. The row of each panel corresponds to a different filter, and each column of a panel corresponds to a different aperture, labeled at the top of each column. Solid lines represent an equal $\Delta m$ between the two methods. Black circles correspond to the saturated sources masked in the SFFT subtraction, while the colored circles represent all other sources.}
    \label{fig:dm_method_comp_sfft}
\end{figure*}

We then decorrelate the difference images output from SFFT using the same procedure as in \citet{HuWang2024a}. This procedure removes correlations induced by the cross-convolution, as well as the SFFT subtraction itself. The decorrelation is performed in tiles across the entire cutout, using a tile with a length 3 times greater than the chosen kernel half-width. After obtaining the decorrelated difference images, we then produce pixel-by-pixel signal-to-noise-ratio (SNR) maps for each cutout following \citet{HuWang2024a}, with 512 Monte Carlo samples. We then mask all pixels on the edges of the decorrelated image and SNR maps, to remove artifacts from the fitting, within a 30 pixel boundary. The image cutouts are overlapped in such a way that all sections of the mosaic are subtracted. Finally, we use the SFFT code to obtain a distribution of SNR values within each cutout.

The procedure used to generate the SFFT difference images and the quality of the subtraction is shown in an example cutout in Fig.~\ref{fig:subtraction_quality}. This example shows that even near bright, saturated stars and diffraction spikes, SFFT can perform well, and is better than direct subtraction. The quality of the SFFT difference images is judged visually, but also by the distribution of the SNR throughout the images. If SFFT models the difference image well, the standard deviation of the SNR pixels belonging to the background should be $\sim$1. Additionally, the lower the standard deviations of the SNR distributions belonging to the sources, the better the model fit. We define an ideal range of standard deviations from background pixels $\sigma_{\rm SNR} \in [0.5, 2.0]$ to achieve in our individual cutout subtractions.

After our initial application of SFFT, we found that nine cutouts with a fraction of non-zero pixels $> 75\%$ in F444W violated our $\sigma_{\rm SNR}$ criterion, overfitting the difference image in most cases. To remedy these specific cutouts, we used a simpler model for the kernel, a first-order B-spline, in order to reduce the flexibility of the model and its tendency to overfit. All other parameters used within SFFT were kept the same, as stated previously. In all cases, this new kernel helped to bring $\sigma_{\rm SNR }$ into the acceptable range. This same procedure was used on the F200W cutouts, and improved a few difference images.

Ultimately, of the 427 (123) cutouts in the F200W (F444W) band, 422 (119) have $\sigma_{\rm SNR}$ that fall into our ideal range, with only 14 (0) cutouts having poor $\sigma_{\rm SNR}$ with a fraction of non-zero pixels $> 75\%$. Poor $\sigma_{\rm SNR}$ can arise from overfitting, usually when there are few sources or non-zero pixels in a cutout, or poor fits, often due to a saturated source occupying most of the cutout.

%\red{Note, to save space, for example figures of multi-panel image stamps, you can combine the direct-sub and SFFT figures, e.g., one left figure and one right figure. }

\subsection{Comparison between Direct Subtraction and SFFT}\label{sec:comp_diffimg}

\begin{figure}
    \centering
    \includegraphics[width=0.5\textwidth]{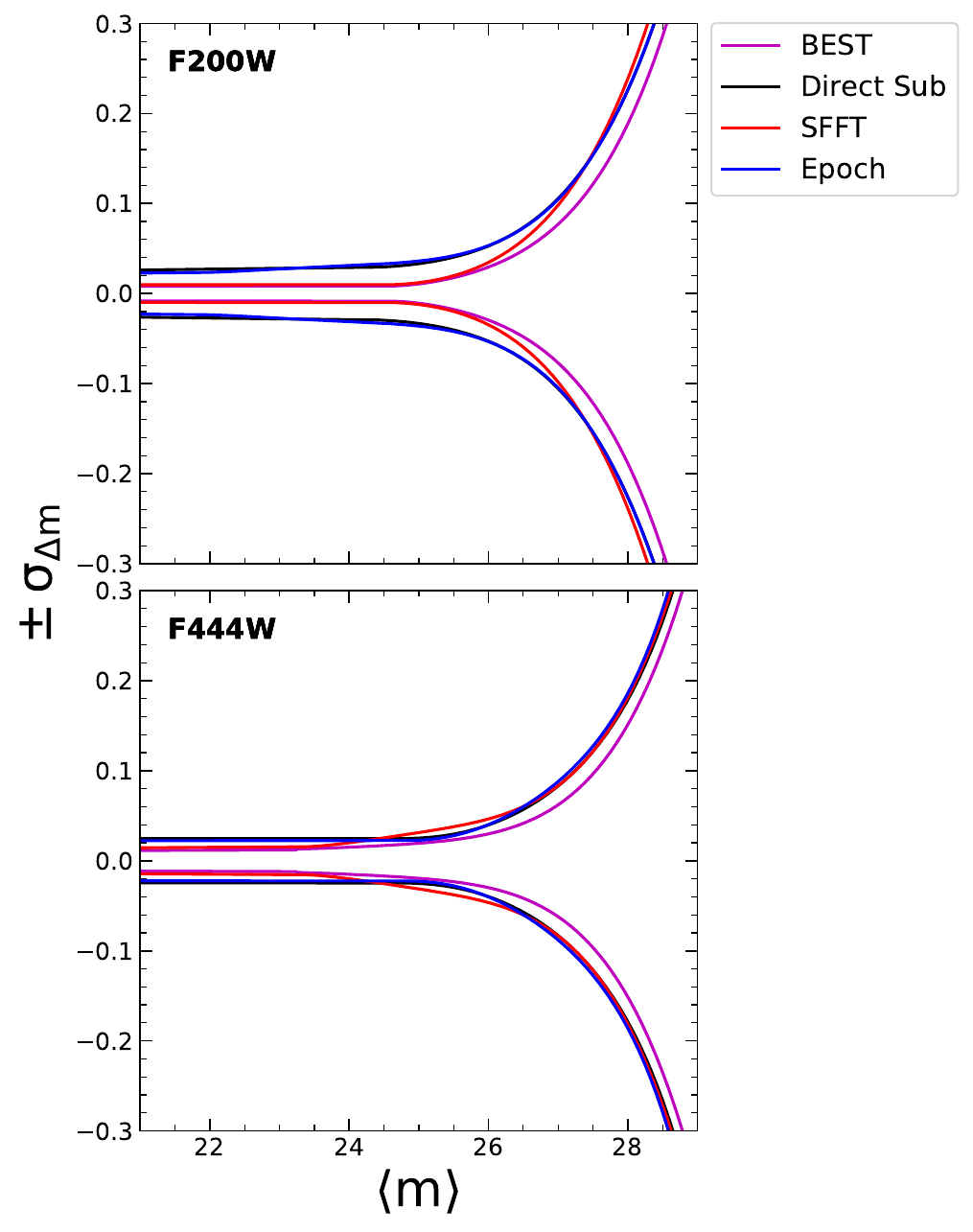}
    \caption{A comparison of the scatter (standard deviation) $\sigma_{\rm \Delta m}$ of the $\Delta m$ distribution as a function of the average source magnitude across two epochs $\langle m \rangle$ between the different methods, using an aperture of 0\farcs2. Each panel shows a different band, displayed in its upper left corner. Each color line represents a different method, as indicated in the upper right legend. The ``BEST'' result is the smaller $\Delta m$ value from either the SFFT or direct-subtraction results. At bright magnitudes ($\langle m\rangle<26$), SFFT produces the best subtraction, while at fainter magnitudes, both SFFT and direct-subtraction can produce better results.  } 
    \label{fig:dm_scatter_method_comp}
\end{figure}

%\red{Here we should demonstrate the improvement, if any, of SFFT over direct subtraction, and show case some winners. }

%\red{NOTE: I hope that SFFT generally performs better than, or at least on par with, direct-sub across the board. But if it is not the case, then one possible solution is to take whichever produces the smallest absolute difference flux for each source. }

% \blue{[ZS]} Generally, SFFT performs slightly better than direct subtraction.

After performing both direct subtraction and SFFT subtraction, we extract the flux from sources across the difference image using {\tt SExtractor} \citep{sextractor} in dual mode, using the same stacked image for detections as in Section \ref{subsec:catalog}. Because the stacked image is defined over the entire NEXUS field using the short-wavelength (SW) channel pixel size, we use {\tt SWarp} \citep{SWarp} to align the stacked image to the difference image in each band. We use similar parameters for this {\tt SExtractor} run as in Section \ref{subsec:catalog}. These difference image sources are then matched to the parent source catalog with a matching radius of 0\farcs1. We further remove all sources that are contaminated by masked pixels within a 0\farcs1 radius. Because the overlapping area between the two epochs depends on the band, there are 23575 (15883) detected sources in the F200W (F444W) band that are covered by both epochs. We show a few example sources and their difference images in Fig.~\ref{fig:example_sources}.

A one-to-one comparison between the SFFT and direct subtraction magnitudes for sources extracted over the whole mosaic is shown in Fig.~\ref{fig:dm_method_comp_sfft}. The two-epoch magnitude difference of these sources $\Delta m$ matches fairly well regardless of band or aperture, with a maximum scatter of $\sim 0.2$ mag from the unity relation. In some cases, SFFT can perform better than direct subtraction if the differential background and source are modeled well. This can be seen for sources where SFFT reveals fainter variability than direct subtraction, causing $|\Delta m_{\rm SFFT}| > 0$ and $|\Delta m_{\rm dir-sub}| \sim 0$. In cutouts where either the source was fit poorly or is a saturated object that was masked in the difference imaging process, artifacts cause direct subtraction to perform better. Additionally, improper background subtraction and small astrometric offsets between the individual epoch imaging can cause spurious detections of variability, where $|\Delta m_{\rm dir-sub}| > 0$ and $|\Delta m_{\rm SFFT}| \sim 0$.

%Further comparisons between these two methods in Fig.~\ref{fig:dm_hist} show that the two methods perform similarly well for all bands and apertures, over the whole sample. 

\begin{figure*}
    \centering
    \includegraphics[width=\linewidth]{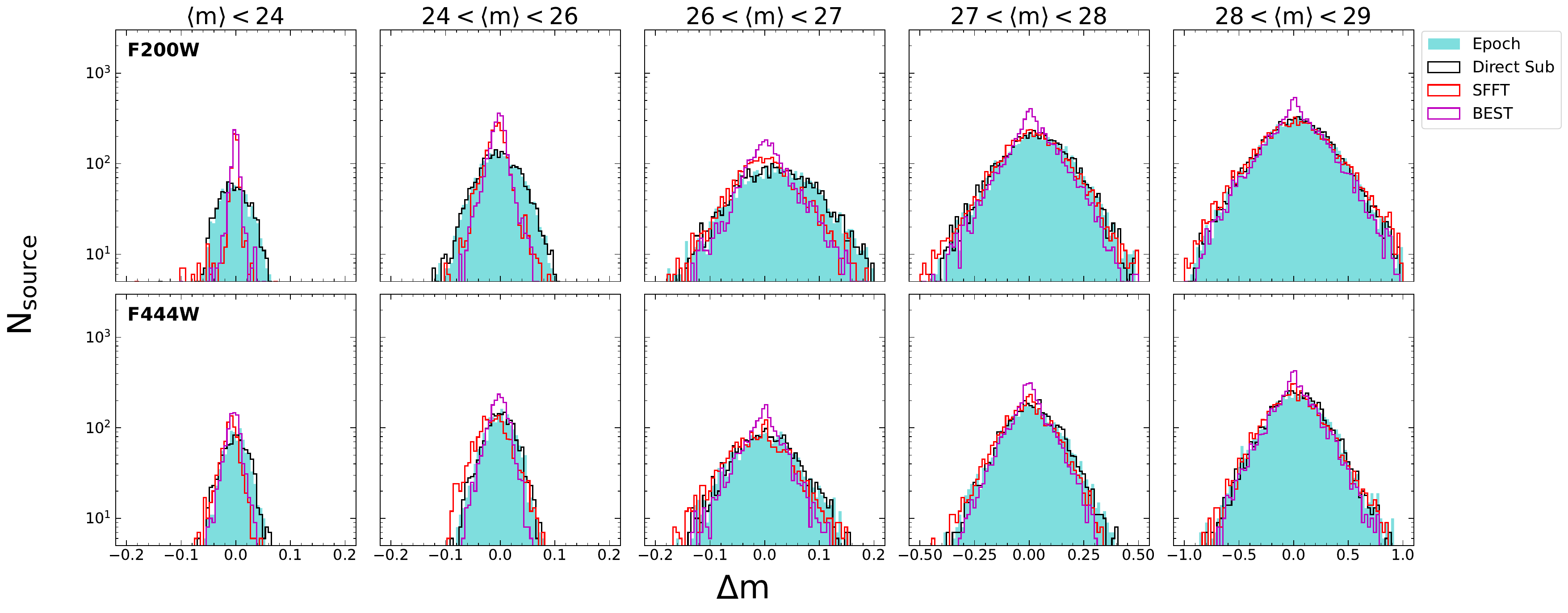}
    \caption{Histograms comparing the distribution of $\Delta m$ for different methods, within different $\langle m \rangle$ bins. Each row represents a unique band, while each column represents a $\langle m \rangle$ bin. Each color represents a different method, identified using the legend in the upper right corner. Note that the spikes at $\Delta m\approx 0$ from difference imaging methods are dominated by well-subtracted sources with difference mag $>35$. These sources are excluded in the measurements of the standard deviation of the $\Delta m$ distribution. }
    \label{fig:dm_scatter_wrt_mavg}
\end{figure*}

To further quantify the improvement of SFFT over direct subtraction, we compare the $\Delta m$ values for all sources in both bands with a 0\farcs2 aperture. We assume the method that performs the best will produce the $\Delta m$ closest to 0. SFFT generally performs better for brighter objects (e.g., Fig.~\ref{fig:dm_scatter_method_comp}), having a lower $|\Delta m|$ for $\sim 80$\% of sources with an average magnitude across epochs $\langle m \rangle < 26$. Direct subtraction performs better for $\sim 80$\% of dimmer objects. SFFT can perform sub-pixel alignment as it matches PSFs across both epochs, making it well-suited to subtracting extended, bright sources. This can fail with small, dim sources, if the PSF alignment performs larger shifts. Such scenarios will produce artifacts near the source, artificially inflating the extracted $\Delta m$. Additionally, if a small source is located in a part of a cutout sparsely covered by other sources, it will be difficult for SFFT to model the spatial variations of the PSF and differential background near that source. Small sources may also not be modeled well by SFFT if they are close to bright, saturated sources.\footnote{A module of SFFT is currently being developed to improve subtraction in such cases involving faint sources, and will be publicly available in a future release.} In these cases, direct subtraction will perform better, as long as the dipole pattern present from sub-pixel misalignment is entirely encompassed by the aperture.

Inspecting the formal uncertainty $\Delta m_{\rm err}$ reported by both methods, the results differ by band and source magnitude. For F200W, the uncertainty from SFFT is generally larger than that from direct subtraction, except for bright sources ($\langle m \rangle < 24$). For F444W, the uncertainty for SFFT is generally smaller than for direct subtraction for all source magnitudes. The higher spatial resolution in F200W causes the fitting in SFFT to be more complex, leading to more artifacts. Dimmer sources contain more artifacts on average, as they are more affected by improper background subtraction and PSF matching.

%\blue{Based on the performance comparison between direct subtraction and SFFT, we take the smaller $\Delta m$ value from the two methods as our fiducial difference imaging results, denoted as $\Delta m_{\rm best}$.  }

\subsection{Improvement over Epoch Photometry}\label{sec:comp_phot}

%Fig.~\ref{fig:dm_method_comp_catalog} compares the magnitude change between the two epochs for sources in the parent catalog using either epoch-photometry or direct subtraction. For all apertures, the epoch-photometry $\Delta m$ correlates well with the direct subtraction $\Delta m$ with minimal scatter, showing consistency across methods. The scatter between $\Delta m_{\rm epoch}$ and $\Delta m_{\rm Direct \; Sub}$ is significantly less than the scatter between $\Delta m_{\rm SFFT}$ and $\Delta m_{\rm Direct \; Sub}$. We expect the magnitude changes from epoch-photometry and direct subtraction to match well if the epochs are aligned well and background is minimal. Investigating this scatter as a function of $\langle m \rangle$, the scatter between $\Delta m_{\rm SFFT}$ and $\Delta m_{\rm Direct \; Sub}$ increases as the source magnitude increases. Contrastingly, the scatter between $\Delta m_{\rm Direct \; Sub}$ and $\Delta m_{\rm Epoch}$ stays relatively constant with respect to source magnitude. Therefore, the majority of the additional scatter, is due to the poorer performance of SFFT with dimmer sources, which make up a significant fraction of our full sample. \red{Based on the explanations of this paragraph, I think we can remove this figure, and go directly to the next figure comparing the $\Delta m$ distributions. }

Fig.~\ref{fig:dm_scatter_method_comp} compares the 3$\sigma$-clipped standard deviation of the $\Delta m$ distribution using different methods as a function of the average source magnitude $\langle m \rangle$, for the default 0\farcs2 aperture case. The scatter from $\Delta m_{\rm SFFT}$ and $\Delta m_{\rm dir-sub}$ increases as the source magnitude increases, as expected from the increased flux uncertainties at faint magnitudes. However, SFFT produces considerably smaller scatter at bright magnitudes ($\langle m\rangle<26$) than both epoch photometry and direct subtraction, and only slightly worse scatter at fainter magnitudes. Combining the SFFT and direct subtraction results, we adopt the smaller $\Delta m$ value from the two methods as our fiducial difference imaging results, denoted as $\Delta m_{\rm BEST}$. This metric is used in our variability analysis below. 

In Fig.~\ref{fig:dm_scatter_wrt_mavg}, we compare the distribution of $\Delta m$ across methods in different source magnitude bins for a 0\farcs2 aperture. At all source magnitudes, epoch photometry overestimates flux variations, especially of dimmer sources. On the other hand, both difference imaging methods produce a significant population of nonvariable sources with $\Delta m \approx 0$. The difference imaging methods are both able to successfully remove the flux background and recover faint source variability. Especially for bright sources in the F200W band, the scatter in $\Delta m$ is significantly lower for either difference imaging method than for epoch photometry. This improvement arises from SFFT's PSF-matching allowing it to properly subtract bright PSF-shaped point sources. Using the fiducial ``best" $\Delta m$ of either difference imaging method causes this decrease in scatter to be much more significant across bands and source magnitude.

%\red{ Fig.~? compares the magnitude change between the two epochs for sources in the parent catalog using different methods: epoch-photometry, direct-subtraction and SFFT difference imaging. Epoch photometry significantly overestimates the flux variations, even with the AUTO aperture. On the other hand, difference imaging methods better subtract the constant source flux, revealing faint variable fluxes. }

\section{Results}\label{sec:results}

%\begin{figure*}
%    \centering
%    \includegraphics[width=\linewidth]{Figures/dm_method_comp_epoch.pdf}
%    \caption{\red{We don't need this figure anymore.} Same as Fig.~\ref{fig:dm_method_comp_sfft}, but the y-axis is $\Delta m$ computed using epoch photometry from the parent catalog.}
%    \label{fig:dm_method_comp_catalog}
%\end{figure*}

% \begin{figure*}
%     \centering
%     \includegraphics[width=\linewidth]{Figures/dm_hist.pdf}
%     \caption{Histograms displaying $\Delta m$ of sources identified in the NEXUS mosaics. The row of each panel corresponds to a different filter, and each column of a panel corresponds to a different aperture, labeled at the top of each column. The black histograms represent $\Delta m$ using direct subtraction, while the colored histograms represent $\Delta m$ using SFFT, discarding masked saturated sources. \red{We probably don't need this figure. }}
%     \label{fig:dm_hist}
% \end{figure*}

\begin{figure*}
    \centering
    \includegraphics[width=\linewidth]{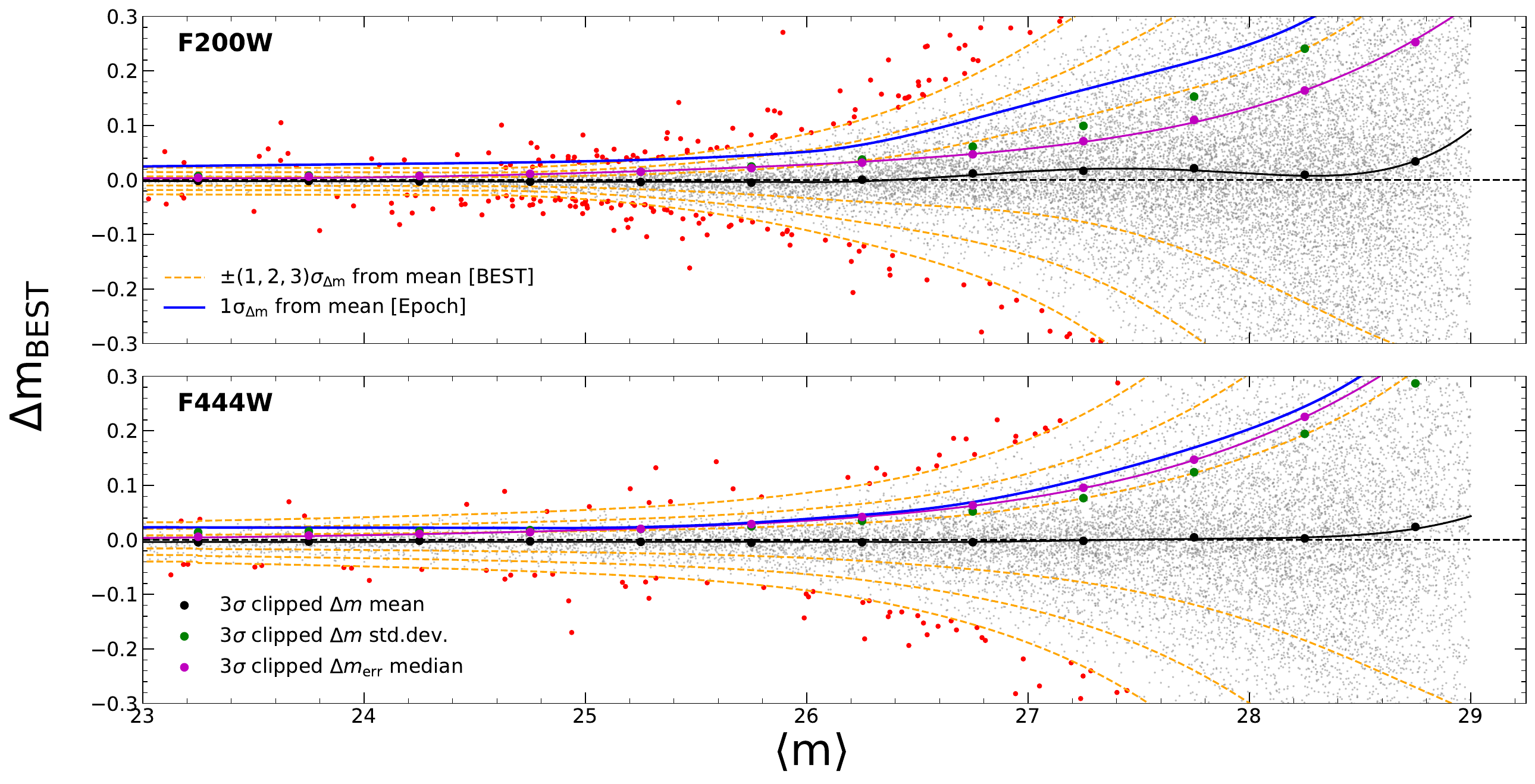}
    \caption{The distribution of $\Delta m_{\rm BEST}$ for all considered bands (rows), using a 0\farcs2 aperture. The 3$\sigma$-clipped, binned ($\Delta m$ mean, $\sigma_{\Delta m}$, $\Delta m_{\rm err}$ median) values are shown using (black, green, magenta) circles. Dashed yellow lines represent the (1,2,3)$\sigma$ deviation from the mean. The solid blue line represents the 1$\sigma$ deviation from the mean using the epoch photometry $\Delta m$. The red points are sources with $|\Delta m_{\rm debiased}| > 3\sigma_{\Delta m}$. Lines fit to the 3$\sigma$-clipped binned ($\Delta m$ mean, $\Delta m_{\rm err}$ median) values are shown in the same color as the binned values. }
    \label{fig:dm_scatter}
\end{figure*}

In Section~\ref{sec:methods} we have demonstrated the significant improvement of difference imaging techniques over epoch photometry in extracting potentially variable fluxes. In particular, the difference imaging techniques produce $\Delta m$ distributions that are significantly narrower and closer to zero than that from the difference in epoch photometry. In other words, difference imaging techniques substantially improve the sensitivity of nuclear variability detection over epoch photometry, thus providing much tighter constraints on the variability (or the lack thereof) measurements. 

%Furthermore, while the SFFT approach performs slightly better than the direct-subtraction approach, there are occasional cases where the SFFT subtraction is not as clean as direct subtraction. Therefore we opt to use the $\Delta m$ value from whichever method (SFFT or direct-sub) that is closer to zero. We designate this $\Delta m_{\rm BEST}$ as our fiducial measurement of variability, and drop its subscript from now on unless necessary. 

%\red{Figures needed: 1. $\Delta m$ comparison between epoch-photometry and difference image -- this figure and associated description below can be moved to the previous section; 2. ensemble variability against mean mag for both direct-sub and sfft methods (in both bands); }

Fig.~\ref{fig:dm_scatter} shows the fiducial $\Delta m_{\rm BEST}$ using either SFFT or direct-subtraction against the source mean mag (between the two epochs) $\langle m \rangle$, for the default 0\farcs2 aperture case. Excluding sources with sufficiently low difference magnitude and those masked as saturated stars for SFFT, we find a nearly symmetric $\Delta m$ distribution with respect to $\langle m \rangle$ for both bands and most apertures. This distribution becomes asymmetric at the faint end of the distribution, due to both the noise and differing depths between the two NEXUS epochs. We find more dimming sources (i.e., $\Delta m > 0$) than brightening ones at large $\langle m \rangle$. This bias in mean $\Delta m$ reaches 0.1 (0.04) mag in F200W (F444W) using a 0\farcs2 aperture at $\langle m \rangle = 29$. Expectedly, the scatter $\sigma_{\Delta m}$ increases with $\langle m \rangle$, and is noticeably larger in F200W than F444W. Comparing between methods, the scatter and asymmetry when using $\Delta m_{\rm BEST}$ are markedly lower than the scatter when using epoch photometry. 

%. The two methods can produce scatters discrepant by a factor of 2--3 at large $\langle m \rangle$, 

To define a threshold for variability detection, we follow the approach in \citet{ZhangZJ2024}, assuming the majority of sources are non-variable. Briefly, we first divide our sample into bins of $\langle m \rangle$ with a width of 0.5 mag. We then calculate the 3$\sigma$-clipped $\Delta m$ standard deviation in each bin, fitting these binned values with a straight line at the bright end and a fourth-order polynomial in the dim end. This smooth $\sigma_{\rm \Delta m}$ curve is used to determine an initial sample of variable sources. Using this curve, we derive the 3$\sigma$-clipped $\Delta m$ and $\Delta m_{\rm err}$ distributions. The latter quantity $\Delta m_{\rm err}$ is the uncertainty in $\Delta m$. The formal uncertainty $\Delta m_{\rm err}$ output by {\tt SExtractor} generally does not represent the true $\Delta m$ uncertainty well, and is thus not used in our variability analysis. Instead, we use the $\sigma_{\Delta m}$ curve to measure the uncertainty floor of $\Delta m$ at each $\langle m\rangle$. We then recover a debiased $\Delta m_{\rm debiased }$, which we use to finalize our sample of variable sources under the condition $|\Delta m_{\rm debiased}| > 3\sigma_{\rm \Delta m}$. 

%We fit these to the same form as $\sigma_{\rm \Delta m}$. 

We show the debiased $\sigma_{\rm \Delta m}$ curves from each method in Fig.~\ref{fig:dm_scatter_method_comp} for the 0\farcs2-diameter aperture. Our fiducial $\Delta m_{\rm BEST}$ produce the lowest scatter among all methods, by a large margin, especially for dim sources. Using the variability criterion defined above and with the best difference imaging $\Delta m_{\rm debiased, BEST}$, we select an initial sample of 735 unique sources that vary significantly in either F200W or F444W using the 0\farcs2 aperture.

\begin{table*}[!t]
    \centering
    \caption{NEXUS Nuclear Variability Source Catalog Format}
    
    \begin{tabular}{|l|l|c|l|}
        \hline
        \textbf{Column Name} & \textbf{Format} & \textbf{Unit} & \textbf{Description} \\
        \hline
        {\tt ID} & LONG & & Unique source ID \\
        {\tt NID} & LONG & & NEXUS ID in EDR, if available \\
        {\tt RA} & DOUBLE & deg & Right Ascension (J2000) \\
        {\tt DEC} & DOUBLE & deg & Declination (J2000) \\
        {\tt Z\_PHOT} & DOUBLE & & $z_{\rm phot}$, if available\\
        {\tt Z\_SPEC} & DOUBLE & & $z_{\rm spec}$, if available\\
        {\tt TWO\_EPOCHS\_[BAND]} & BOOL & & Whether the object was observed for both epochs \\
        {\tt SATURATED\_[BAND]} & BOOL & & Whether the source was masked as saturated for SFFT \\
        \hline
        \tablenotemark{$*$}{\tt MAG\_AVG\_[BAND]} & DOUBLE(4) & mag & $\langle m \rangle$ \\
        \tablenotemark{$*$}{\tt FLUX\_[BAND]\_[IMAGE]} & DOUBLE(4) & $\mu$Jy & Source flux \\
        \tablenotemark{$*$}{\tt FLUXERR\_[BAND]\_[IMAGE]} & DOUBLE(4) & $\mu$Jy & Source flux uncertainty \\
        \tablenotemark{$*$}{\tt DM\_RAW\_[BAND]\_[METHOD]} & DOUBLE(4) & mag & $\Delta m$ \\
        \tablenotemark{$*$}{\tt DMERR\_RAW\_[BAND]\_[METHOD]} & DOUBLE(4) & mag & $\Delta m_{\rm err}$ \\
        \tablenotemark{$*$}{\tt SIGMA\_DM\_[BAND]\_[METHOD]} & DOUBLE(4) & mag & $\sigma_{\Delta m}$ at the source's $\langle m \rangle$ \\
        \tablenotemark{$*$}{\tt DM\_DEBIASED\_[BAND]\_[METHOD]} & DOUBLE(4) & mag & $\Delta m_{\rm debiased}$ \\
        \tablenotemark{$*$}{\tt DMERR\_DEBIASED\_[BAND]\_[METHOD]} & DOUBLE(4) & mag & Corrected $\Delta m_{\rm err}$  \\
        \tablenotemark{$*$}{\tt VARIABLE\_[BAND]\_[METHOD]} & BOOL(4) & & If the source has $|\Delta m_{\rm debiased}|>3\sigma_{\Delta m}$ \\
        \hline
        {\tt REF\_FLUX\_[BAND]} & DOUBLE & $\mu$Jy & Fiducial REF source flux \\
        {\tt SCI\_FLUX\_[BAND]} & DOUBLE & $\mu$Jy & Fiducial SCI source flux \\
        {\tt DIFF\_FLUX\_[BAND]} & DOUBLE & $\mu$Jy & Fiducial ``BEST" source flux \\
        {\tt REF\_FLUXERR\_[BAND]} & DOUBLE & $\mu$Jy & Fiducial REF source flux uncertainty \\
        {\tt SCI\_FLUXERR\_[BAND]} & DOUBLE & $\mu$Jy & Fiducial SCI source flux uncertainty \\
        {\tt DIFF\_FLUXERR\_[BAND]} & DOUBLE & $\mu$Jy & Fiducial ``BEST" source flux uncertainty \\
        {\tt DM\_RAW\_[BAND]} & DOUBLE & mag & Fiducial $\Delta m$ \\
        {\tt DMERR\_RAW\_[BAND]} & DOUBLE & mag & Fiducial $\Delta m_{\rm err}$ \\
        {\tt SIGMA\_DM\_[BAND]} & DOUBLE & mag & {Fiducial $\sigma_{\Delta m}$} \\
        {\tt DM\_DEBIASED\_[BAND]} & DOUBLE & mag & {Fiducial $\Delta m_{\rm debiased}$ in 0\farcs2-diameter aperture} \\
        {\tt DMERR\_DEBIASED\_[BAND]} & DOUBLE & mag & Fiducial corrected $\Delta m_{\rm err}$ \\
        {\tt VARIABLE\_[BAND]} & BOOL & & Fiducial variability flag \\
        {\tt MASK\_VISUAL\_INSPECTION\_[BAND]} & BOOL & & Whether the source was visually inspected as variable \\
        \hline
        {\tt ELONGATION\_[BAND]\_[TYPE]} & DOUBLE & & The {\tt SExtractor} ELONGATION parameter \\
        {\tt FLAGS\_[BAND]\_[IMAGE]} & LONG & & The {\tt SExtractor} FLAGS parameter \\
        {\tt IMAFLAGS\_ISO\_[BAND]\_[IMAGE]} & LONG & & The {\tt SExtractor} IMAFLAGS\_ISO parameter \\
        {\tt CLASS\_STAR\_[BAND]\_[IMAGE]} & DOUBLE & & The {\tt SExtractor} CLASS\_STAR parameter \\
        {\tt ISOAREA\_IMAGE\_[BAND]\_[IMAGE]} & LONG & pix & The {\tt SExtractor} ISOAREA\_IMAGE parameter \\
        \hline
    \end{tabular}
    \vspace{5pt}
    { \footnotesize
    \begin{adjustwidth}{0cm}{}
    Fiducial variability candidates can be selected with {\tt VARIABLE\_[BAND]=True} \& {\tt MASK\_VISUAL\_INSPECTION\_[BAND]=True}.\\[5pt] 
        \tablenotemark{
        $*$}These columns have four values for each source, one for each aperture. The apertures used are in the following order: AUTO, 0\farcs2, 0\farcs3, 0\farcs5. \\[5pt]
        Columns denoted with {\tt[BAND]} are two columns, one for each band (F200W or F444W). \\[5pt]
        Columns denoted with {\tt[TYPE]} are two columns, one for each type of image (DIFF or EPOCH). DIFF refers to the difference images, while EPOCH refers to the two NEXUS epoch mosaics. \\[5pt]
        Columns denoted with {\tt[METHOD]} are four columns, one for each method of difference imaging (SFFT, DIRECTSUB, EPOCH, or BEST). \\[5pt]
        Columns denoted with {\tt[IMAGE]} are five columns, one for each image used (REF, SCI, SFFT, DIRECTSUB, or BEST). Here, REF is the reference image (Wide Epoch 1) and SCI is the science image (Deep Epoch 1). 
    \end{adjustwidth}
    }
    \label{tab:catalog}
\end{table*}

% \clearpage

We carefully examine the identified variable sources that pass our criterion above, 
i.e., $|\Delta m_{\rm debiased}|>3\sigma_{\Delta m}$. Upon visual inspection, we flag 36\% of these variable sources as being potentially artificial, which are caused by systematics in the image subtraction. We flag sources if their difference images have significant artifacts near the source, and such is the case for bright and point-source sources. Such artifacts arise from poor PSF-matching, poor astrometric alignment, and improper background model fitting.\footnote{Future releases of SFFT will describe such artifacts and offer ways to mitigate them.} Variable sources with dipole artifacts from image misalignment that are sufficiently contained within the aperture are kept within the sample, as the structure from the dipole will not bias the aperture flux measurement. We also remove objects close to the edges of the mosaics, or near bad pixels, which may affect nearby pixels. If a variable object, say, contains an artifact in F200W, but not in F444W, we leave it in the sample, but remove its variability flag in F200W. The majority of the remaining 465 variables are robust and further discussed in Section~\ref{sec:disc}. We compile all variability measurements for our publicly available\footnote{\url{https://ariel.astro.illinois.edu/nexus/paper_data/nuclear_variability/nexus_wide01_deep01_stacked_sources_allband.fits.gz}} source catalog in Table~\ref{tab:catalog}. Variable sources that pass our visual inspection have {\tt MASK\_VISUAL\_INSPECTION\_[BAND]= True} in this table.

As a sanity check, we verify that our difference imaging approach is able to recover transient candidates identified by the NEXUS transient search team via an independent effort, as long as these transients are present in the stacked image and included in the parent catalog as distinct sources. Such isolated transients typically show significant detection in the difference flux, and are often undetected in one of the two epochs. However, our specific variability criterion is designed for nuclear variability, where the persistent source is well detected in both epochs with a well-defined average magnitude. For this reason, some of the transients with detection in only one epoch may not pass our formal variability threshold using an ill-defined average source magnitude. Nevertheless, our difference flux measurements (with formal uncertainties) are retained in Table~\ref{tab:catalog} to select such transient candidates. For example, one can select such transient variables using {\tt DIFF\_FLUX\_[BAND]}$>5\times${\tt DIFF\_FLUXERR\_[BAND]} for 5$\sigma$ detections in the difference flux. Notably, our difference imaging approach can recover both brightening and dimming transients. 

% One example transient detection is shown in Fig.~\ref{fig:variable_sources} (last row).

\section{Discussion}\label{sec:disc}

\begin{figure}
    \centering
    \includegraphics[width=0.48\textwidth]{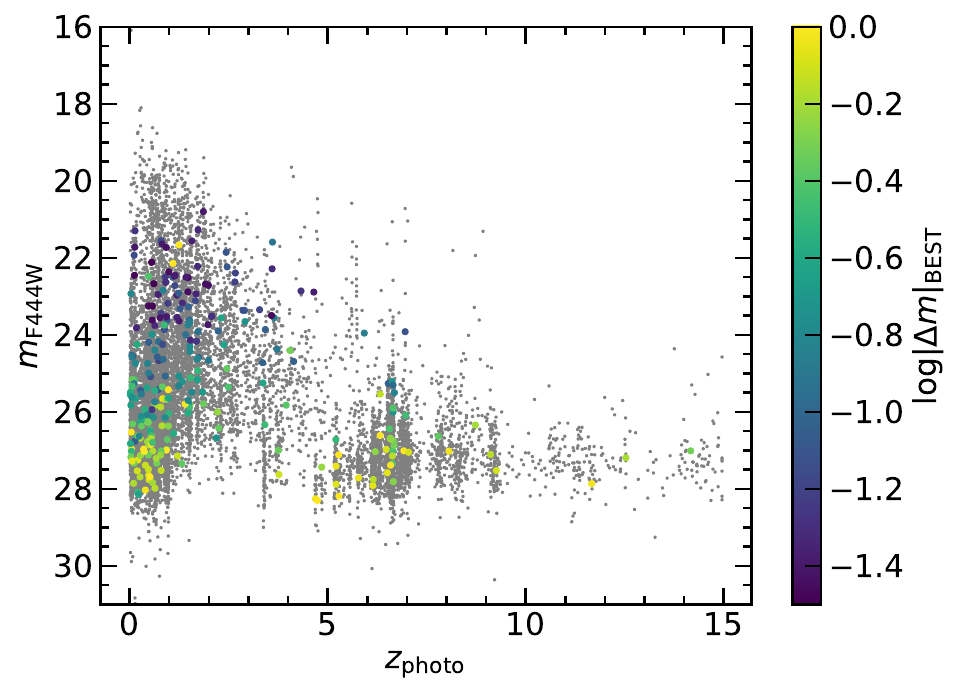}
    \caption{Distribution of the sample of high-confidence nuclear variables in the photometric redshift - $m_{\rm F444W}$ plane, color-coded by the maximum variability in the two bands. These nuclear variables follow a similar photo-z distribution as the underlying sample (gray points). }
    \label{fig:zdist}
\end{figure}

We now proceed to discuss the nature of the variable sources detected by our difference imaging process. Our fiducial 0\farcs2 diameter aperture ensures the detected variable flux is within 1~kpc from the nucleus of the source at all redshifts. We use the morphological classification {\tt CLASS\_STAR} and a preliminary version of photometric redshift $z_{\rm phot}$ catalog for NEXUS sources to separate extragalactic sources from Galactic stars. Our photometric redshift estimation procedure is described in \citet{ZhuangEtAl2025a}. Among the $\sim 450$ visually vetted variable sources, we attempt to classify stars with {\tt CLASS\_STAR}$>0.9$ in both bands and $z_{\rm phot}<0.1$, and find that all reliable variables are extragalactic. The distribution of these nuclear variables in the photometric redshift versus F444W magnitude plane is shown in Fig.~\ref{fig:zdist}. These variable sources follow a similar redshift distribution as the parent NEXUS galaxy sample. 

% Briefly, we utilize {\tt EAZY} \citep{EAZY} with NIRCam photometry in all six filters and accompanying Subaru Hyper Suprime-Cam photometry.

%\red{Plot $|\Delta m_{f444w}|_{best}$ vs $|\Delta m_{f200w}|_{best}$ for the subset of variable sources that have 2-epoch coverage in both bands? For this plot, perhaps restrict to sources with $\langle m \rangle<26$. Note that the error floor is pretty flat below 26, then it rises as the flux uncertainty increases due to the imaging depth. }

%In Fig.~? we show the distribution of nuclear distance of the variable sources for the fiducial 0\farcs2 aperture case. \red{This figure will ensure us that most of the variable sources are nuclear. }

%\red{We may divide this section into subsections depending on the results. E.g., LRD variability might be a separate subsection. }

A detailed classification of the variable population requires spectroscopy, which is currently scarce. NEXUS has been accumulating NIRCam/WFSS and NIRSpec/MSA spectroscopy for sources in the field, and we will perform detailed studies of these variable sources in future work. Below, we showcase several examples with nuclear variability identified from this work, for which we have reliable spectroscopic classifications from WFSS or MSA spectra, or reasonable source classifications based on NIRCam photometry from the NEXUS EDR \citep{nexus-edr}. As of now, we mostly have classifications for AGNs, LRDs, and off-nuclear transients (see \citealt{nexus-qdr}), but as the NEXUS survey progresses, our source classification will extend to various types of astrophysical sources (e.g., TDEs, variable stars, etc.).

%\red{Discussion items:
%\begin{enumerate}
%    \item discuss the nature of these significant variables, such as their source morphology, the location of the variable source (nuclear or not), the photo-z distribution, and any confirmed spectral types of these variable sources; any LRDs in the variable sources (defer to Sec~5.2 for LRDs); mention variable sources that are only detected in one epoch -- these are extreme variables with very large $\Delta m$, and might be transients
%    \item discuss the general variability properties -- for the variable population, describe the limit that our observations can place; 
%\end{enumerate}}

\subsection{Case Studies}

%Here we present and discuss special cases of interest, such as robust nuclear flares, extreme variables, confirmed brown dwarfs, etc.

%\red{Here we present and discuss special cases of interest, such as robust nuclear flares, extreme variables, confirmed brown dwarfs, etc. Show their detection images. We should also show their 2-band light curves. See if we spot any important cases that are worth follow-up studies (not in this paper). }

% \red{Here let's comment on the nature of each individual of the examples shown in Fig.~\ref{fig:variable_sources}. Let's show 5 examples? }

\begin{figure*}
    \centering
    \includegraphics[width=0.8\linewidth]{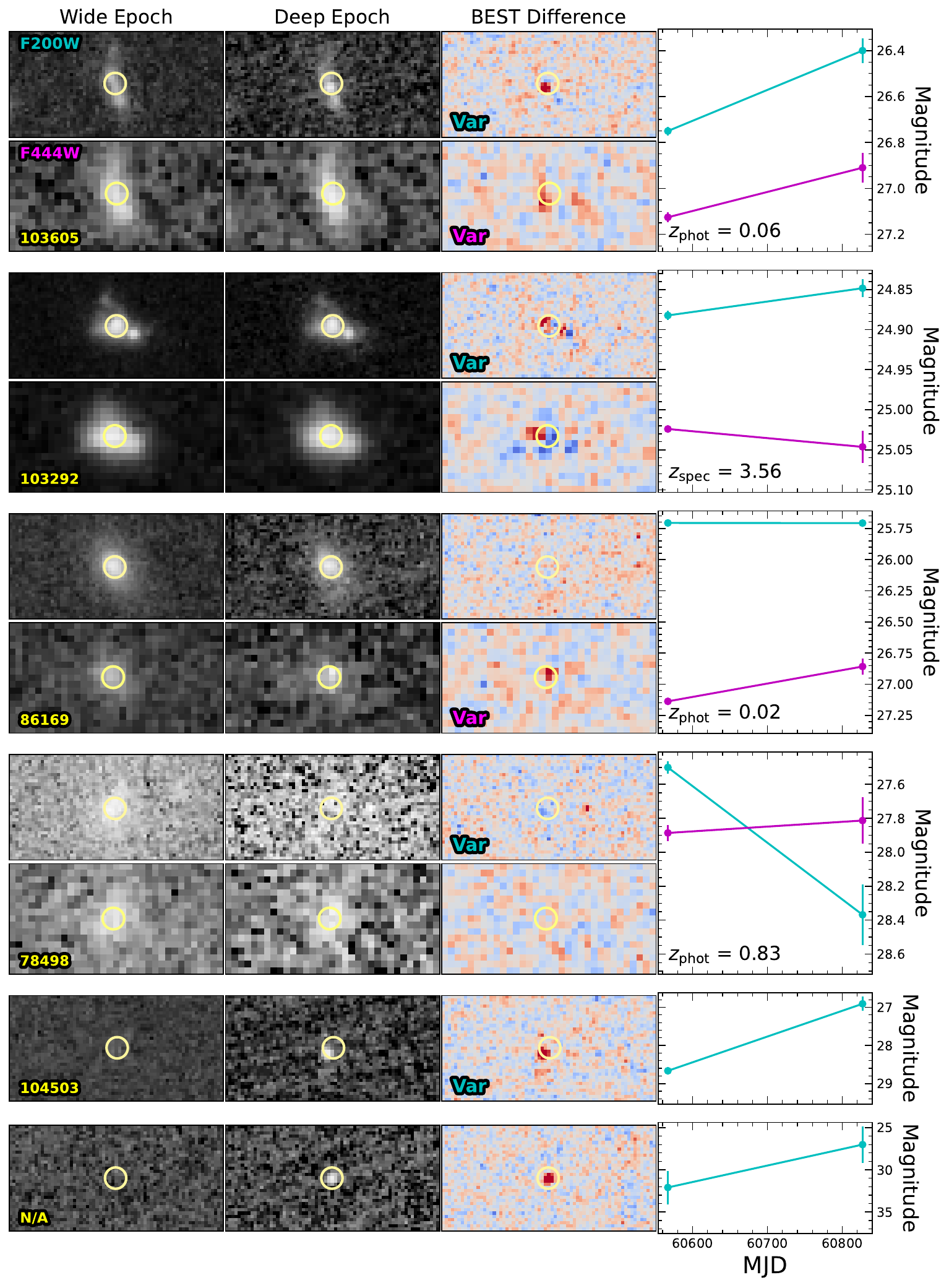}
    \caption{A few example source cutouts (2\arcsec$\times$1\arcsec) with significant variability detected by our approach. For each source, each column corresponds to the reference, science or ``BEST" difference image. The difference image uses a linear color scale, where blue is negative, red is positive, and the limits of the color map are $\pm 5\sigma$ of the background pixels. A 0\farcs2 diameter aperture is drawn around the source in each image in red. The top (bottom) row for each source corresponds to the F200W (F444W) band. The source ID (not NID) is shown in the lower left panel. The rightmost column shows the light curve for each source in each band, where cyan (magenta) represents F200W (F444W). The Deep epoch magnitude is obtained using the Wide epoch magnitude and $\Delta m_{\rm debiased}$. We use $\sigma_{\Delta m}$ as the uncertainty in $\Delta m_{\rm debiased}$. If available, the $z_{\rm spec}$ or $z_{\rm phot}$ of the object is shown the lower left of the rightmost panel. If a source was visually inspected as truly variable in a given band, it is labeled in the bottom left of the difference image. The final two sources only have coverage in F200W, so F444W is not shown.}
    \label{fig:variable_sources}
\end{figure*}

We showcase a few examples with significant nuclear variability detected by difference imaging in Fig.~\ref{fig:variable_sources}. These objects were chosen due to their significant variability, unique multi-band variability, and significance as candidates of certain astrophysical transients. Whenever possible, we list the spectroscopic or photometric redshift of the source. The majority of the variable candidates we identified and inspected are likely normal broad-line AGNs, where the F200W variability is more significant than the variability in F444W. But there are cases where the F444W variability is more significant than that in F200W, making the interpretation more complicated. Nuclear transients or tidal disruption events with more extinction in F200W than F444W may cause their variability differences. Below, we elaborate on these individual cases. 

Source 103605 is a nearby ($z_{\rm phot} = 0.06$) AGN candidate showing significant nuclear variability in both bands, with $\Delta m_{\rm debiased} = -0.35\ (-0.21)$ in F200W (F444W), with a low $\sigma_{\Delta m}$. There is a clear bright spot shown in the difference image in both bands, as well. Similar sources are desirable for future AGN variability analysis, as multi-band variability can be used to grant tighter constraints.

Source 103292 is a relatively bright, high-redshift ($z_{\rm spec} = 3.56$) spectroscopically monitored AGN candidate, displaying significant variability in F200W, with $\Delta m_{\rm debiased} = -0.03$. Two nuclei can be seen in the NEXUS epoch mosaics, and both nuclei display variability in the difference images in both bands (albeit with some residuals). The spectroscopy for this source shows evidence for \halpha\, and \SIII\, emission. Additionally, it possibly dims in F444W but brightens in F200W. Variability within AGNs can be correlated between bands, usually with a certain characteristic time delay resulting from properties of the SMBH and gas surrounding it \citep{Peterson2001}. Even if correlated, if the delay between bands is long enough, $\sim$months in this case, it can brighten in one band and dim in another. This seemingly dual-AGN system will be useful for constraining the evolution of dual SMBHs in the early Universe.

Not many sources in our sample vary more in F444W than in F200W. Source 86169 ($z_{\rm phot} = 0.02$), however, does vary significantly in F444W ($\Delta m_{\rm debiased} = -0.28$) with minimal $\sigma_{\Delta m}$, but not in F200W. The variability in source 86169 is slightly off-nucleus, suggesting this may be a supernova or other off-nuclear transient.

Our variability detection procedure not only recovers brightening sources, but also dimming ones. Source 78498 ($z_{\rm phot} = 0.83$) is detected in F200W in the Wide epoch, and seemingly vanishes in the Deep epoch, dimming by $\Delta m_{\rm debiased} = 0.86$. Such seemingly uncorrelated multi-band variability may indicate that this source hosts an AGN like source 103292. Interestingly, another brightening off-nuclear transient can be seen near this object, $\sim$0\farcs5 to the right of the nucleus. 

There are a few sources in our sample that demonstrate extreme nuclear variability (e.g., $|\Delta m_{\rm debiased}| > 1$). Source 104503 is within both epoch mosaics only for F200W, but varies extremely ($\Delta m_{\rm debiased} = -1.76$), much larger than $\sigma_{\Delta m}$. While source 104503 does not have a reliable $z_{\rm phot}$, it is worth following up in future NEXUS epochs. This source is point-like, suggesting a possible supernova or nuclear flare.

We also investigate the transient candidates identified by the NEXUS transient team but not formally included in our parent source catalog. The final row in Fig.~\ref{fig:variable_sources} shows a hostless transient (RA=268.4957263, DEC=65.22676717), only seen in F200W. We utilize {\tt photutils} \citep{photutils} to obtain an estimate of the 0\farcs2 aperture flux of this transient in both epochs and the difference images. We use the same bias curve and $\sigma_{\Delta m}$ curve as for the rest of our sample to de-bias $\Delta m$ and obtain an estimate of $\sigma_{\Delta m}$. This bright transient is clearly identifiable in the difference image in F200W, with $\Delta m_{\rm debiased} = -5.08$, certainly passing our variability criterion. Many of the other dozens of NEXUS transients are also distinctly variable in our difference images. Future NEXUS epochs will help to identify more hostless transients and constrain their true nature.

\subsection{LRD Variability}

We examine the variability properties of the 10 spectroscopically confirmed LRDs at $3\lesssim z\lesssim 7$ from the NEXUS EDR \citep{ZhuangEtAl2025a} that are covered by both epochs. We focus on spectroscopically confirmed LRDs because photometric samples of LRDs generally contain substantial contamination from non-broad-line AGNs and are therefore nonvariable. 

\begin{figure}
    \centering
    \includegraphics[width=0.48\textwidth]{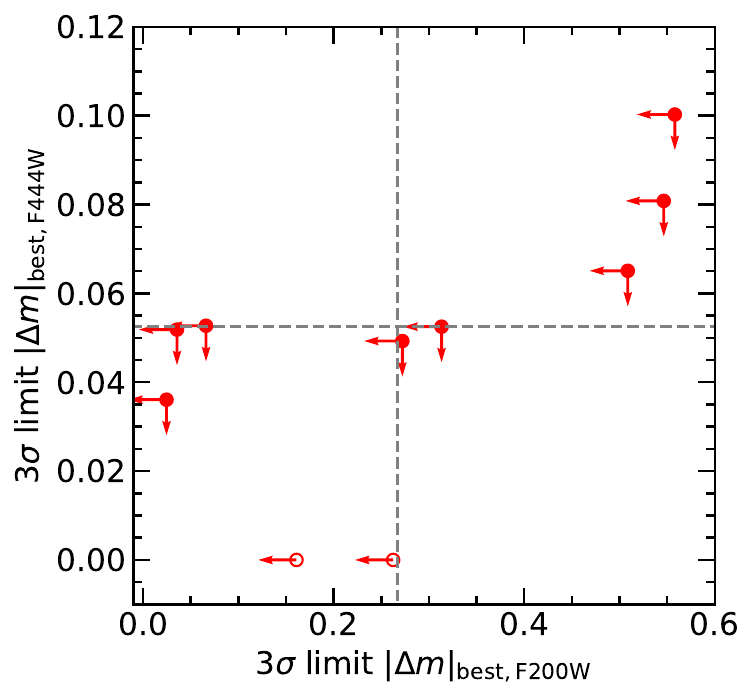}
    \caption{$3\sigma$ upper limit of variability for the 10 spectroscopically-confirmed LRDs in \citet{ZhuangEtAl2025a} based on our variability analysis. Two objects do not have F444W variability measurements and are marked as open circles at the bottom. The dashed lines mark the median values of the 3$\sigma$ upper limits. } 
    \label{fig:lrd_var}
\end{figure}

For the 10 spectroscopic LRDs, none of them have detected variability in either band from our variability analysis. Fig.~\ref{fig:lrd_var} shows the 3$\sigma$ upper limit of variability for the 10 LRDs. The variability constraints in F444W flux, which traces the rest-frame optical of these LRDs, are stringent: the $3\sigma$ upper limit ranges between $\sim 3-10\%$ with a median value of $\sim 5\%$. These constraints are consistent with those reported by \citet{Kokubo2024} and \citet{ZhangEtAl2025}, though our constraints are somewhat tighter. This weak variability in the rest-frame optical of LRDs provides important constraints on the BH accretion nature of these objects and motivates continued theoretical investigations \citep[e.g.,][]{LiuEtAl2025, SecundaEtAl2025}. For example, some recent theoretical models of LRDs suggest super-Eddington accretion onto the central SMBH and that the rest-frame optical continuum comes from thermal emission from a photosphere with an effective temperature of $\sim 5000~$K \citep{LiuEtAl2025}. Such models generally predict weak continuum variability undetectable by NEXUS \citep[e.g.,][]{SecundaEtAl2025}. On the other hand, the variability constraints in F200W are less stringent, given the larger flux uncertainties in the SW band. \citet{ZhuangEtAl2025a} suggest that the LRD flux in F200W (rest-frame UV) is spatially extended over tens of parsecs, and hence is not originated from nuclear accretion. Even if the extended rest-frame UV emission is from scattered AGN light, the light-crossing time across tens of parsecs of the reflector would result in a non-detection of variability over the timescales probed by NEXUS observations. We therefore do not expect to detect variability in F200W for these LRDs. 

%Here we can test if the photometric LRD sample show notably more variability. Then, we focus on the few spectroscopically confirmed broad-line LRDs with variability measurements. 

\section{Conclusions}\label{sec:con}

In this work, we perform a systematic search for nuclear variability in F200W+F444W among the $\sim 25,000$ sources in the NEXUS field that have two NIRCam imaging epochs between September 2024 and June 2025. We test the performance of epoch photometry and of two different image subtraction techniques: direct subtraction and SFFT. We find a significant improvement of the sensitivity of variability detection using difference imaging over epoch photometry. Our best performance reaches $1-2\%$ relative photometry for $m<25$ sources, and $\sim 0.1-0.2$~mag at $m\sim 28$. 

We identify 465 reliable variable sources that lie above $3\sigma$ of our best difference imaging sensitivity as a function of source brightness and show no obvious image subtraction systematics upon visual inspection. Essentially, all these variability candidates are of extragalactic origin. While spectroscopic confirmation of their nature is still scarce, the vast majority of these variable sources are likely broad-line AGNs, with a minority of them likely from unidentified nuclear transients, such as high-redshift tidal disruption events and supernovae.

We also investigate the variability of LRDs. For the 10 spectroscopically confirmed broad-line LRDs reported in \citet{ZhuangEtAl2025a}, none of them show detectable variability in either band. We derive $3\sigma$ upper limits on the F444W variability of $\sim 3-10\%$ for these LRDs, with a median value of $\sim 5\%$. This weak variability in rest-frame optical of LRDs is consistent with earlier findings \citep[e.g.,][]{Kokubo2024,ZhangEtAl2025}. 

As the NEXUS program continues to monitor the field with NIRCam imaging at a regular 2-month cadence through 2028, we will be able to compile light curves for all variable sources therein and improve variability measurements. In addition, the subset of nuclear variables with $m_{\rm F444W}<26$ identified from our approach will supply important targets for the NIRSpec/MSA spectroscopic follow-up in NEXUS \citep{nexus}. In turn, these spectra will help constrain the nature and properties of these bright variables. However, the majority of these nuclear variables are too faint for the NEXUS NIRSpec/MSA observations, and require deep NIRSpec spectroscopy from dedicated JWST programs.

\begin{acknowledgements}
The JWST data presented in this article were obtained from the Mikulski Archive for Space Telescopes (MAST) at the Space Telescope Science Institute. The specific observations analyzed can be accessed via \dataset[doi: 10.17909/azms-tb18]{https://doi.org/10.17909/azms-tb18}. This work utilizes resources supported by the National Science Foundation's Major Research Instrumentation program, grant \#1725729, as well as the University of Illinois at Urbana-Champaign. JDRP is supported by NASA through an Einstein Fellowship grant No. HF2-51541.001 awarded by the Space Telescope Science Institute (STScI), which is operated by the Association of Universities for Research in Astronomy, Inc., for NASA, under contract NAS5-26555. L.H. acknowledges support from the STScI grant JWST-AR-05965. Based on observations with the NASA/ESA/CSA James Webb Space Telescope obtained from the Barbara A. Mikulski Archive at the Space Telescope Science Institute, which is operated by the Association of Universities for Research in Astronomy, Incorporated, under NASA contract NAS5-03127. Support for Program numbers JWST-GO-05105 was provided through a grant from the STScI under NASA contract NAS5-03127.
\end{acknowledgements}

\software{
\texttt{Astropy} \citep{2013A&A...558A..33A,2018AJ....156..123A,astropy_3},
\texttt{EAZY} \citep{EAZY},
\texttt{Matplotlib} \citep{Hunter2007}, 
\texttt{NoiseChisel} \citep{noisechisel},
\texttt{Numpy} \citep{Harris2020}, 
\texttt{photutils} \citep{photutils}, 
\texttt{scipy} \citep{scipy},
\texttt{SExtractor} \citep{sextractor},
\texttt{SFFT} \citep{HuEtAl2022a},
\texttt{STPSF} \citep{webbpsf},
\texttt{SWarp} \citep{SWarp}.
}

% \appendix

% \begin{figure}
%     \centering
%     \includegraphics[width=0.5\linewidth]{artifacts.png}
%     \caption{Caption}
%     \label{fig:placeholder}
% \end{figure}

%% Appendix material should be preceded with a single \appendix command.
%% There should be a \section command for each appendix. Mark appendix
%% subsections with the same markup you use in the main body of the paper.

%% Each Appendix (indicated with \section) will be lettered A, B, C, etc.
%% The equation counter will reset when it encounters the \appendix
%% command and will number appendix equations (A1), (A2), etc. The
%% Figure and Table counter will not reset.

\bibliography{sample631, ZRefs}{}
\bibliographystyle{aasjournal}

%% This command is needed to show the entire author+affiliation list when
%% the collaboration and author truncation commands are used.  It has to
%% go at the end of the manuscript.
%\allauthors

%% Include this line if you are using the \added, \replaced, \deleted
%% commands to see a summary list of all changes at the end of the article.
%\listofchanges

\end{document}